\documentclass[12pt]{article}

\usepackage{graphicx}

\newcommand{\lsim}{\raisebox{-0.13cm}{~\shortstack{$<$ \\[-0.07cm] $\sim$}}~}
\newcommand{\gsim}{\raisebox{-0.13cm}{~\shortstack{$>$ \\[-0.07cm] $\sim$}}~}

\newcommand{\imag}{\Im {\rm m}}
\newcommand{\real}{\Re {\rm e}}

\begin{document}

\mbox{ } \\[-1cm]
\mbox{ }\hfill KIAS-03077       \\
\mbox{ }\hfill KEK-TH-866       \\
\mbox{ }\hfill UAB-FT-542       \\
\mbox{ }\hfill MC-TH-2003-2     \\
\mbox{ }\hfill KAIST-TH 2003/02 \\
\mbox{ }\hfill hep-ph/0401024    
\bigskip
\thispagestyle{empty}
\setcounter{page}{0}
\begin{center}
{\Large{\bf
  Threshold corrections to $m_b$ and the $b\bar{b}\to H^0_i$   \\[3mm]
   production in CP-violating SUSY scenarios}}                 \\[5mm]
 Francesca Borzumati$^1$, Jae Sik Lee$^2$, and Wan Young Song$^3$
\end{center}
\begin{center}
$^1$ {\it KIAS, 207-43 Cheongryangri 2-dong, Dongdaemun-gu,      
       Seoul 130-722, Korea}                                   \\[1mm]
$^2${\it 
Department of Physics ans Astronomy, University of Manchester, \\
Manchester M13 9PL, United Kingdom}                            \\[1 mm]
$^3${\it 
  Department of Physics, KAIST, Daejeon 305-701, Korea}
\end{center}
\vskip 1.5cm

\begin{abstract}
The inclusion of supersymmetric threshold corrections to the $b$-quark
mass has dramatic consequences in scenarios with large CP-mixing
effects in the Higgs sector. In particular, when the phase of the
combination $M_{\tilde{g}}\mu$ is~$\sim 180^{\rm o} \pm 30^{\rm o}$,
the lightest sbottom squark becomes tachyonic and, possibly, the
$b$-quark Yukawa coupling nonperturbative for values of $\tan \beta$
ranging from intermediate up to large or very large, depending on the
size of $arg(A_t\mu)$, $arg(A_b\mu)$, and the details of the spectrum.
For these scenarios, when allowed, as well as scenarios with different
values of $arg(M_{\tilde{g}}\mu)$, the cross sections for the
production of the three neutral Higgs bosons through $b$-quark fusion 
have interesting dependences on $arg(A_t\mu)$ and $arg(A_b\mu)$, and
the deviations induced by the $m_b$ corrections are rather large.  In
general, such production channel cannot be neglected with respect to
the production through gluon fusion.  For large CP-mixing effects, the
lightest neutral Higgs boson can be mainly CP odd and the $b$-quark
fusion becomes its main production mechanism. Searches at the Tevatron
and the LHC can easily detect such a Higgs boson, or constrain the
CP-violating scenarios that allow it.
\end{abstract}

\newpage

\setlength{\parskip}{1.01ex} 

It is well known that threshold corrections to the $b$-quark mass due
to the virtual exchange of supersymmetric particles can be large and
do not decouple in the limit of heavy superpartners~\cite{bMASScorr}.
They turn out to be quite substantial~\cite{DEMIR}, also for moderate
values of $\tan \beta$, in scenarios that maximize the CP-mixing
effects induced in the Higgs sector at the quantum level by phases in
the supersymmetric and soft supersymmetry-breaking mass parameters
(such scenarios were first studied in
Refs.~\cite{CPVHiggs0,CPVHiggs}).

These corrections affect the vertex $\bar{b}$-$b$-$H^0_i$, which is
responsible for one of the production mechanisms of neutral Higgs
bosons at hadron colliders. If no CP-mixing effects are present in the
Higgs sector, $H^0_i$ is either one of the two CP-even states $h$ and
$H$, or the CP-odd state $A$~\cite{REVIEWS}. In this case, the
mechanism of $b$-quark fusion has relevance, for example, for the
production of $A$~\cite{Higgsrev} (and in the decoupling limit, also
in the case of the production of $H$).  Indeed, a factor 
$\tan \beta$ in the $\bar{b}$-$b$-$A$ coupling can give a production
rate comparable to, or even dominant over, the rate of production
through gluon fusion.  On the contrary, lacking this $\tan \beta$
factor, the $b$-quark fusion plays a minor role in the production of
the light CP-even Higgs boson with respect to the gluon-fusion, which,
although loop-induced, is advantaged by the large yield of gluons in
the proton.  In scenarios with a CP-violating Higgs sector, the three
mass eigenstates $H^0_i$ 
are mixed states with both CP-even and CP-odd 
components~\footnote{Phenomenological consequences of these mixings can 
be found in Refs.~\cite{EXP-CP-L,EXP-CP-H,EXP-CP-OTHER}.}. Thus,
it is conceivable that, if the CP-violating mixing effects are large,
the production through $b$-quark fusion may be relevant for all mass
eigenstates of neutral Higgs bosons.

The aim of this paper is to investigate this issue, while consistently
including the supersymmetric threshold corrections to the $b$-quark
mass.  We find that these corrections can have a strong impact if
phases are present in supersymmetric and supersymmetry-breaking
parameters, in that they can exclude intermediate/large values of
$\tan \beta$, leaving nevertheless allowed the very large ones.
Predictably, when these corrections and the CP-mixing effects in the
Higgs sector are large, the production cross sections through
$b$-quark fusion for all the neutral Higgs bosons are affected, with
that for the lightest one, possibly, dramatically enhanced. 
The effect of the supersymmetric
threshold corrections to $m_b$ on these cross sections 
may remain rather large also when some phases in these scenarios 
acquire trivial values, i.e. $0^{\rm o}$ or $180^{\rm o}$.

We illustrate our findings for scenarios that have been dubbed CPX
scenarios~\cite{CPX}, and that tend to maximize the CP-mixing effects
in the Higgs sector~\footnote{In general, large CP-violating phases in
 supersymmetric models are indirectly forbidden by the non-observation
 of electron and neutron electric dipole moments.  These constraints,
 however, can be evaded by cancellations between the one- and
 higher-loop contributions~\cite{EDMcontrib} to the electric dipole
 moments if the first two generations of sfermions are heavier than
 ${\cal O}(1\,$TeV)~\cite{EDMlargephases,EDMandCOUPLINGS,EDMconstr}. 
 Although somewhat tuned, these scenarios have sparked quite some
 interest, and need to be probed directly through collider searches of
 CP violation in the Higgs sector.}. These scenarios are
identified by the spectrum:
\begin{equation}
|A_t|=|A_b| = 2 c_A \,M_{\rm SUSY}\,, 
\quad
|\mu|=        4 c_{\mu} \,M_{\rm SUSY}\,,
\quad
m_{{\widetilde Q}_3,{\widetilde U}_3,{\widetilde D}_3} =M_{\rm SUSY}\,, 
\quad
|M_{\widetilde g}|=1\,{\rm TeV}\,,
\label{eq:CPXpara}
\end{equation}
where $A_t$ and $A_b$ are the complex trilinear soft terms for the
third generation squarks, $\mu$ is the complex supersymmetric
Higgs(ino) mass parameter; $m_{{\widetilde Q}_3}$, 
$m_{{\widetilde U}_3}$ and $m_{{\widetilde D}_3}$ are the real
soft-breaking masses for the third generation squarks; 
$M_{\widetilde{g}}$ is the complex gluino mass, and $c_A, c_{\mu}$ 
are real numbers.  (The mass parameters entering the slepton sector
and the mass of the two weak gauginos are irrelevant for our
discussion.)  The remaining parameter needed to specify the Higgs
sector is chosen here to be the pole mass of the
charged Higgs boson, $m_{H^\pm}$~\footnote{We remind that the issue of
 electroweak-symmetry breaking, possibly radiatively induced as in the
 constrained Minimal Supersymmetric Standard Model is not addressed in
 these scenarios.}.  Strictly speaking the spectrum of CPX scenarios
has $c_A=c_{\mu}=1$. Here, we extract informations also for cases with
$c_A, c_{\mu}<1$.  We take $M_{\rm SUSY}=0.5\,$TeV; vary $\tan\beta$,
the ratio of the {\it v.e.v.}'s $v_d$ and $v_u$ acquired by $H_d^0$
and $H_u^0$ at the minimum of the Higgs potential, and vary also
$m_{H^\pm}$. As for the phases~\cite{CPMSSM} present in these
scenarios, we take as free parameters the arguments of the products
$A_t\mu$, $A_b\mu$, which we assume to be equal, and of the product
$M_{\tilde{g}}\mu$, respectively 
$\Phi_{A\mu}\equiv{\rm arg}(A_t\mu)={\rm arg}(A_b\mu)$, and 
$\Phi_{g\mu}\equiv{\rm arg}(M_{\tilde{g}}\mu)$.  When relevant, we 
discuss also the {\it four} CP-conserving cases obtained for
$\Phi_{A\mu},\Phi_{g\mu}=0^{\rm o}, 180^{\rm o}$.  Our numerical
analyses make use of the recently-developed program
CPsuperH~\cite{CPsuperH}.

We start observing that, once the threshold corrections to the
$b$-quark mass are included, the in-general-complex $b$-quark Yukawa
coupling is
\begin{equation}
 h_b = \frac{\sqrt{2}\,m_b}{v\cos\beta}\,\frac{1}{R_b}\,,
\label{eq:hb}
\end{equation}
where $v^2  = v_d^2 + v_u^2$, with $v\simeq 254\,$GeV, and 
$v_d = v \cos \!\beta$, $v_u = v \sin \!\beta$. The factor $ R_b$:
\begin{equation}
 R_b=1+\kappa_b\tan\beta
\end{equation}
collects in $\kappa_b$ the finite corrections~\footnote{There are
additional corrections $\delta h_b/h_b$ to $R_b$ which are not
 enhanced by $\tan\beta$. We neglect them in our discussion but their
 effect is fully included in the numerical analysis~\cite{CPsuperH}.}
to the $b$-quark mass. In turn, $\kappa_b$ can be split as
\begin{equation}
 \kappa_b=\epsilon_g+\epsilon_H\,,
\end{equation}
where $\epsilon_g$ and $\epsilon_H$ are, respectively, the
contribution coming from the sbottom-gluino exchange diagram and from
the stop-Higgsino diagram. Their explicit form is
\begin{equation}
\epsilon_g=
    \frac{2\alpha_s}{3\pi} \, M_{\widetilde g}^* \mu^*
    I(m_{\tilde{b}_1}^2,m_{\tilde{b}_2}^2,|M_{\widetilde g}|^2)\,, 
\quad \quad
\epsilon_H=
    \frac{|h_t|^2}{16\pi^2} \, A_t^* \mu^*
    I(m_{\tilde{t}_1}^2,m_{\tilde{t}_2}^2,|\mu|^2)\,.
\label{eq:epsilon}
\end{equation}
The one-loop function $I(a,b,c)$ is well known and can be found, for 
example, in~\cite{BFPT}. The left-right mixing elements in the 
matrices for the sbottom- and stop-mass squared used to obtain these 
corrections, are, in our convention,  
\begin{equation}
\frac{1}{\sqrt{2}} \, h_b^* (A_b^*  v_d - \mu v_u)\,, \quad \quad 
\frac{1}{\sqrt{2}} \, h_t^* (A_t^* v_u - \mu v_d)\,.  
\end{equation}
For the spectrum of Eq.~(\ref{eq:CPXpara}) with $c_A=c_\mu=1$ and
$M_{\rm SUSY}=0.5$\,TeV, it is 
$|\mu|^2, |M_{\widetilde g}|^2 \gg M^2_{\rm SUSY}$ and, as a
consequence, the sbottom-gluino corrections are, in absolute value,
much larger than those obtained from stop-Higgsino exchange.  For
$\alpha_s $ and $h_t$ at the scale $M_{\rm SUSY}$, i.e.  
$\alpha_s \sim 0.1$ and $h_t \sim 1$, $|\epsilon_g|$ has a value 
$\sim 0.05$ and it is about one order of magnitude larger than
$|\epsilon_H|$ ($|\epsilon_g|$ is larger than $|\epsilon_H|$ by the
factor
$\sim \pi\,|\mu|^2/|A_t M_{\widetilde g}|$)~\footnote{Such large
 values of $|\epsilon_g|$ and $|\epsilon_H|$ are specific to the
 scenarios considered here and are compatible with those found in
 Ref.~\cite{IbNa}.}.  In this case, the threshold corrections to the
$b$-quark Yukawa coupling are strongly affected by the phase
$\Phi_{g\mu}$, whereas the dependence on $\Phi_{A\mu}$ is weak.  We
remind that, at the one-loop level, the corrections inducing the
CP-mixing in the Higgs sector are sensitive only to $\Phi_{A\mu}$.
$\Phi_{g\mu}$ affects this mixing at the two-loop level through the 
one-loop corrections to the $b$- and $t$-quark masses~\cite{CPVHiggs}.
Notice that the hierarchy $|\epsilon_g| \gg |\epsilon_H|$ holds in
general, except in the somewhat unlikely cases 
$|M_{\widetilde g}| \ll |A_t|$ or 
$|\mu^2|\,,M^2_{\rm SUSY} \ll |A_tM_{\widetilde g}|$. Possible
variations of the renormalization scale of $\alpha_s$ and $h_t$ do not
affect this statement, neither have any substantial impact on the
numerical results for the production cross sections of the various
states $H_i$.

The effect of the radiative corrections becomes most significant for
$\Phi_{g\mu}=\Phi_{A\mu}=180^{\rm o}$, when dramatic constraints on
$\tan\beta$ can be obtained. Indeed, to prevent the lighter
sbottom squark $\tilde{b}_1$ from becoming tachyonic, the 
$b$-quark Yukawa coupling is constrained as
\begin{equation}
 |h_b| \, \lsim \, 
 \frac{\sqrt{2}\,M_{\rm SUSY}^2}{v\,|\mu|}
\left[1+{\cal O}\!\left(\frac{|A_b|}{|\mu|\tan\beta}\right)\right]\,,
\label{eq:hbexclusion}
\end{equation}
where we have taken $v_u \sim v$ and neglected the $m_b^2$- and
$D$-term contributions to the diagonal elements of the matrix for the
sbottom-mass squared.  Since ${\cal O}({|A_b|}/{|\mu|\tan\beta})$ is
in general negligible, the constraint on $h_b$ can be easily recast
into a constraint on $\tan\beta$. That is, the region of $\tan\beta$:
\begin{equation}
\frac{1}{|\kappa_b|+\displaystyle{\frac{m_b|\mu|}{M_{\rm SUSY}^2}}}
\ \lsim \ \tan\beta 
\ \lsim \
\frac{1}{|\kappa_b|-\displaystyle{\frac{m_b|\mu|}{M_{\rm SUSY}^2}}}
\label{eq:tanbexclusion}
\end{equation}
is {\it not} allowed when
\begin{equation}
 |\kappa_b| > \frac{m_b|\mu|}{M_{\rm SUSY}^2}\,, 
\quad \quad \quad 
 \kappa_b   = -|\kappa_b|\,.
\label{eq:kbcondition}
\end{equation}
These conditions are easily satisfied by the spectrum in
Eq.~(\ref{eq:CPXpara}), for any value of $c_A$ and $c_{\mu}$.

\begin{figure}[t]
\begin{center}
\includegraphics[width=7.3cm]{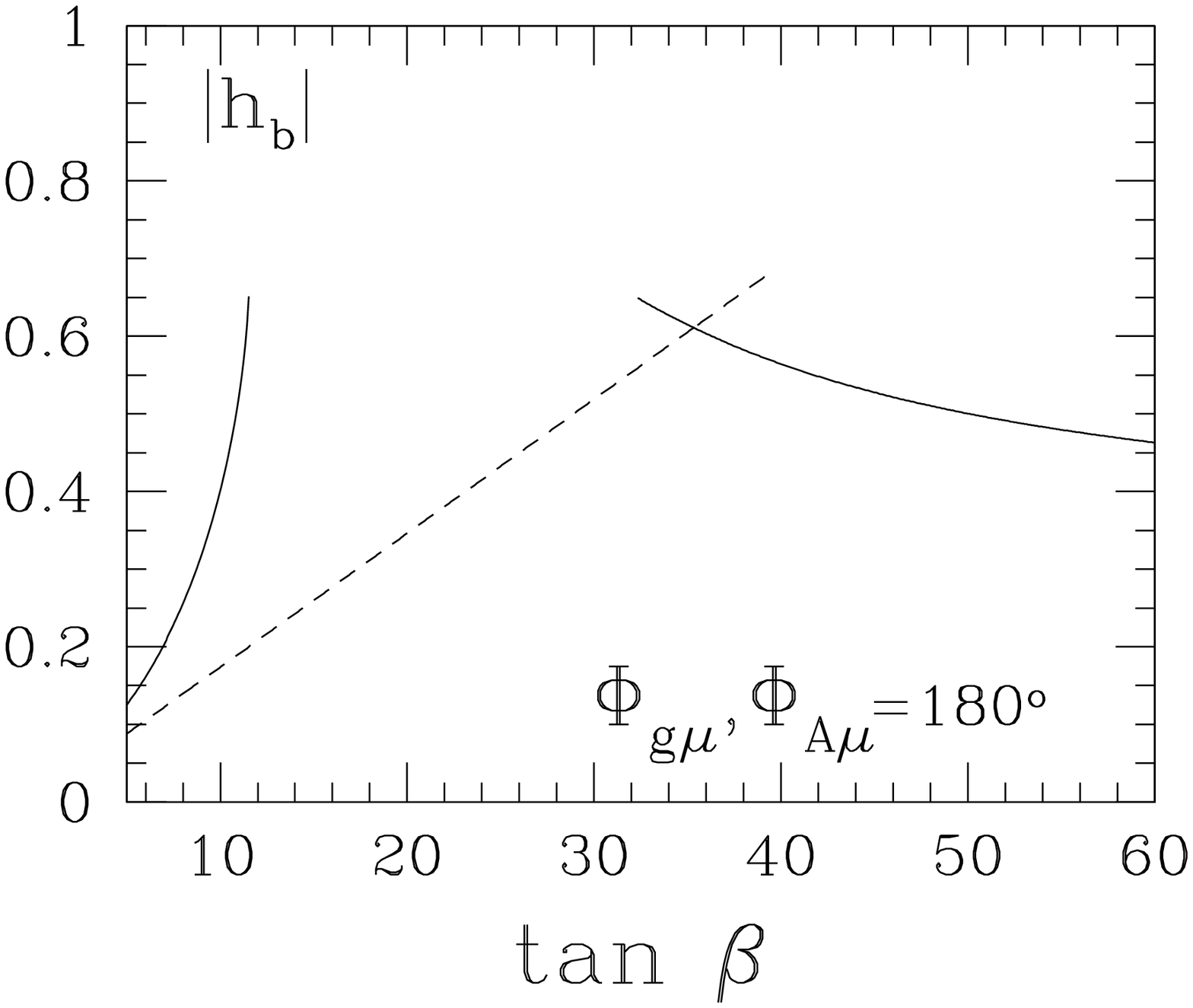}
\hspace{5mm}
\includegraphics[width=7.3cm]{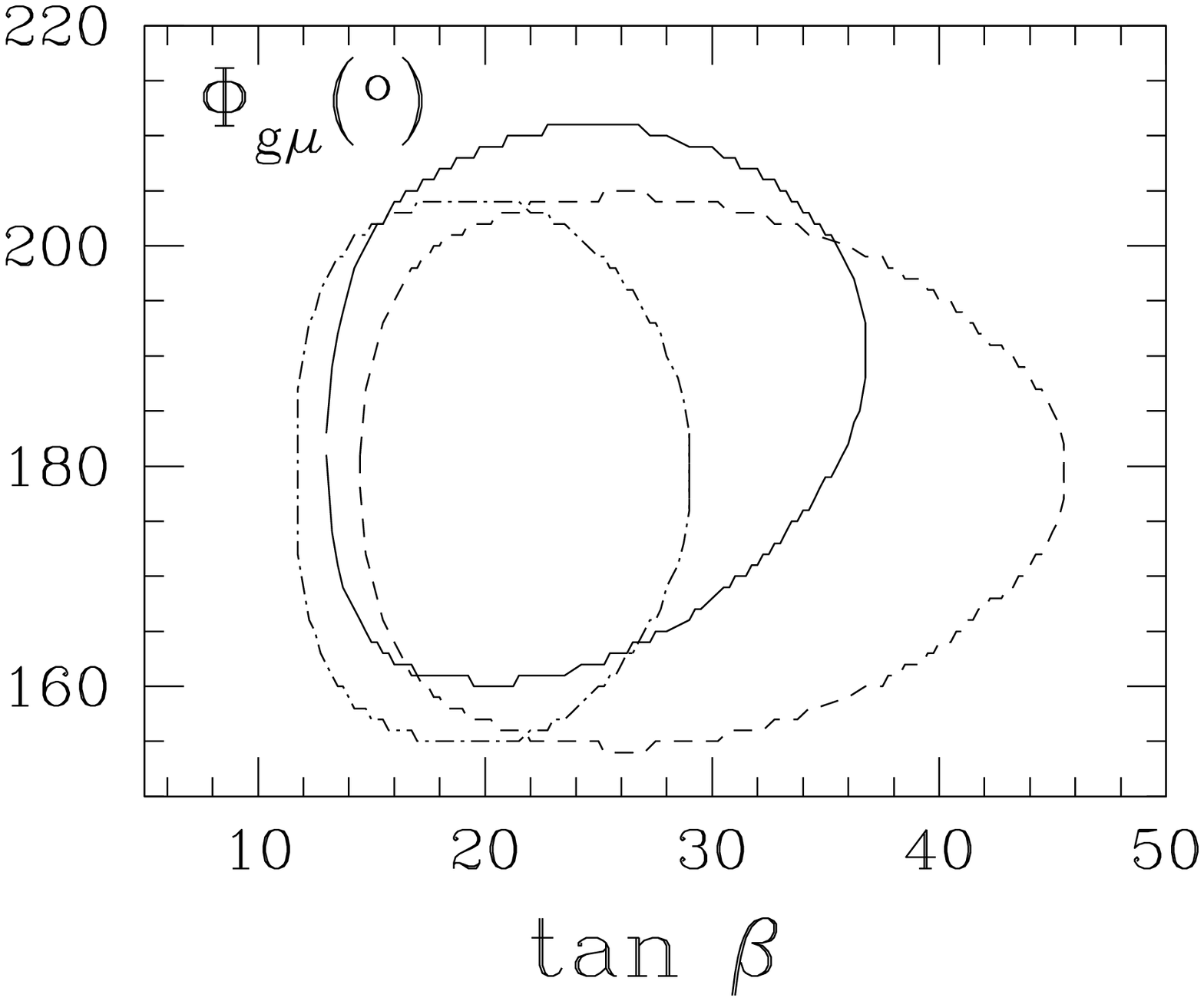}
\end{center}
\vspace{-0.4cm}
\caption{{\small \it Absolute value of the Yukawa coupling of the 
 $b$ quark, vs. $\tan \beta$ (left frame) for the scenario of
 Eq.~\ref{eq:CPXpara}, with $c_A=c_{\mu}=1$, $M_{\rm SUSY} =0.5\,$TeV,
 and $\Phi_{A\mu}=\Phi_{g\mu}=180^o$. For the same scenario, but 
 fixed values of $\Phi_{A\mu}$, the regions of the plane
 $(\Phi_{g\mu},\tan\beta)$ in which the mass of the
 $\tilde{b}_1$ squark is vanishing are shown on the right frame.  They
 are enclosed by a dashed, solid, dot-dashed line corresponding to
 $\Phi_{A\mu}=0^{\rm o},90^{\rm o},180^{\rm o}$.}}
\label{fig:exclusion}
\end{figure}

The constraint of a nontachyonic $\tilde{b}_1$ excludes the 
$\tan \beta$ region $12\lsim\tan\beta\lsim 30$ for the scenario in
Eq.~(\ref{eq:CPXpara}) with $M_{\rm SUSY}=0.5\,$TeV, $c_A =c_{\mu}=1$,
and $\Phi_{g\mu}=\Phi_{A\mu}=180^{\rm o}$. The value $m_b=3\,$GeV was
used here and will be used also for all other numerical evaluations
throughout this paper.  In such a case, the coupling $h_b$ has at most
the value $0.7$ in the allowed region, as shown by the left frame of
Fig.~\ref{fig:exclusion}.  By imposing that the lightest-Higgs boson
mass exceeds, say, $115\,$GeV~\footnote{This value is reminiscent of
 the lower bound on the lightest-Higgs boson from
 LEP2~\cite{LEP2}. This was, however, obtained in the CP-conserving
 case and in the limit of a heavy CP-odd Higgs boson.  Moreover, it is
 known~\cite{CPX} that in CP-violating scenarios the lower bound that
 can be deduced from LEP data may be considerably weaker than that
 reported in Ref.~\cite{LEP2}.  In our case case, the
 choice of this value is motivated by the fact that it allows simple
 comparisons with existing results in the literature. Particularly
 important will be later on the comparison with results presented in
 Ref.~\cite{CHL}, where the same choice is made.},
the excluded region gets extended up to $\tan\beta\sim 50$.  Notice
that the limiting case $\tan \beta \to\infty$, i.e. $ v_d \to 0$, is
not a problematic one, when it comes to obtain acceptable values of
$m_{\tilde{b}_1}$ and $m_{H_1}$.  The $h_b$ coupling is, in this case,
$h_b \sim \sqrt{2} m_b /v \kappa_b$ and the mass of the $b$ quark is
generated at the quantum level~\cite{BFPT}.  For smaller values of
$c_\mu$, for example such that $|\mu| = M_{\rm SUSY}$, the excluded
region of $\tan \beta$ at intermediate/large values is further
enlarged by the fact that $|h_b|$ grows rapidly above perturbative
levels.

This $\tan\beta$ exclusion occurs also for nontrivial values of
$\Phi_{g\mu}$ and $\Phi_{A\mu}$, as it can be seen in the right frame
of Fig.~\ref{fig:exclusion}. In it, we show explicitly the regions of
the plane $\Phi_{g\mu}$-$\tan\beta$ in which the mass of the
$\tilde{b}_1$ squark is negative.  They are enclosed by a dashed,
solid, dot-dashed line, corresponding to
$\Phi_{A\mu}=0^{\rm o},90^{\rm o},180^{\rm o}$, respectively.  The
scenario chosen for this frame is, again, that of
Eq.~(\ref{eq:CPXpara}) with $M_{\rm SUSY}=0.5\,$TeV and 
$c_A=c_{\mu}=1$.  Regions of $\tan \beta$ to be excluded 
are found for $\Phi_{g\mu} \sim 180^{\rm o} \pm 30^{\rm o}$. 
This pattern has to be compared with the very large
$\tan \beta$ exclusion typical of constrained supersymmetric models
with $\Phi_{A\mu}=\Phi_{g\mu}=0^{\rm o}$, in which the threshold
corrections to $m_b$ are not taken into account~\cite{BOP}.

\begin{figure}[p]
\begin{center}
\includegraphics[width=7.3cm]{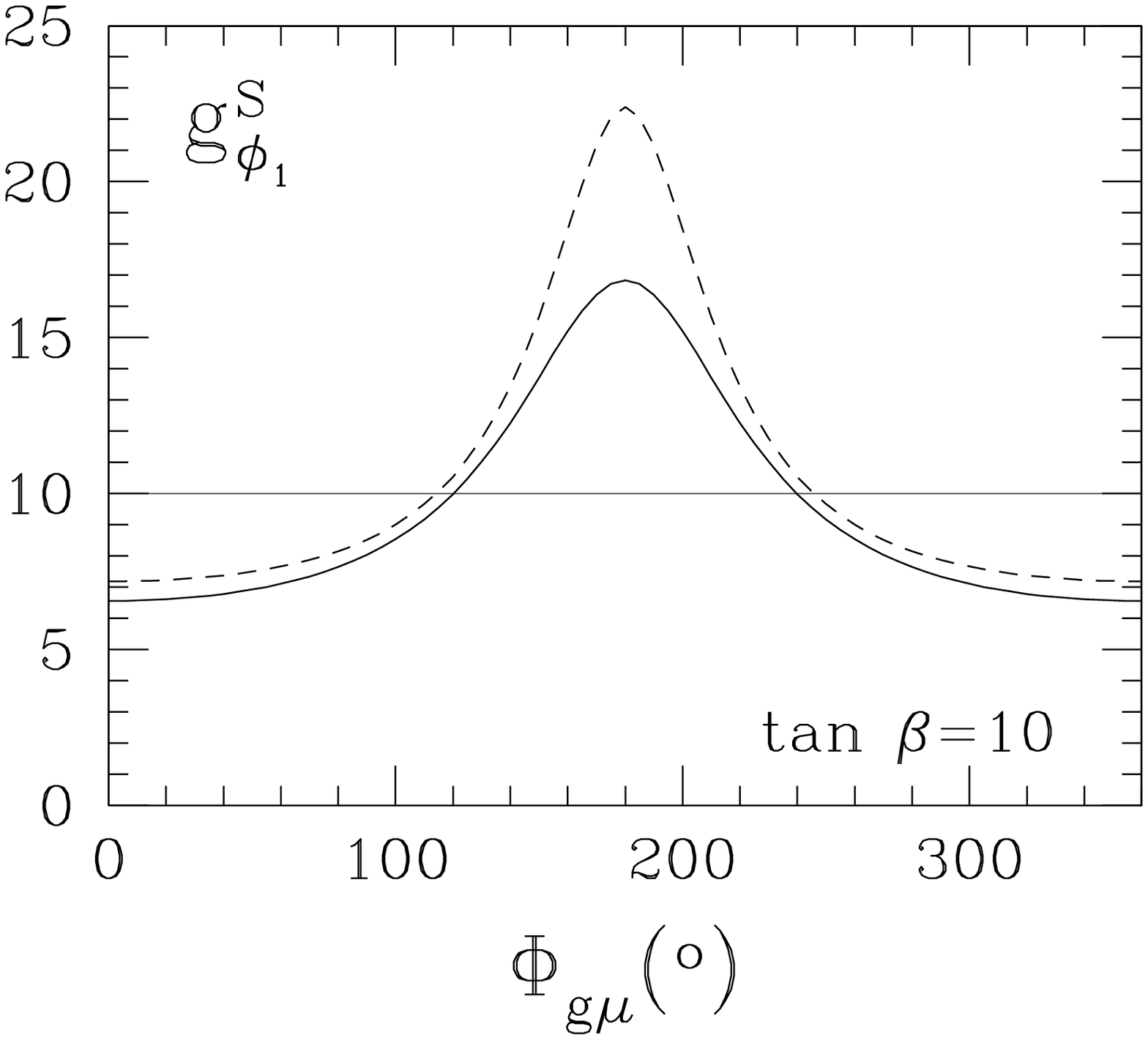}
\hspace{5mm}
\includegraphics[width=7.3cm]{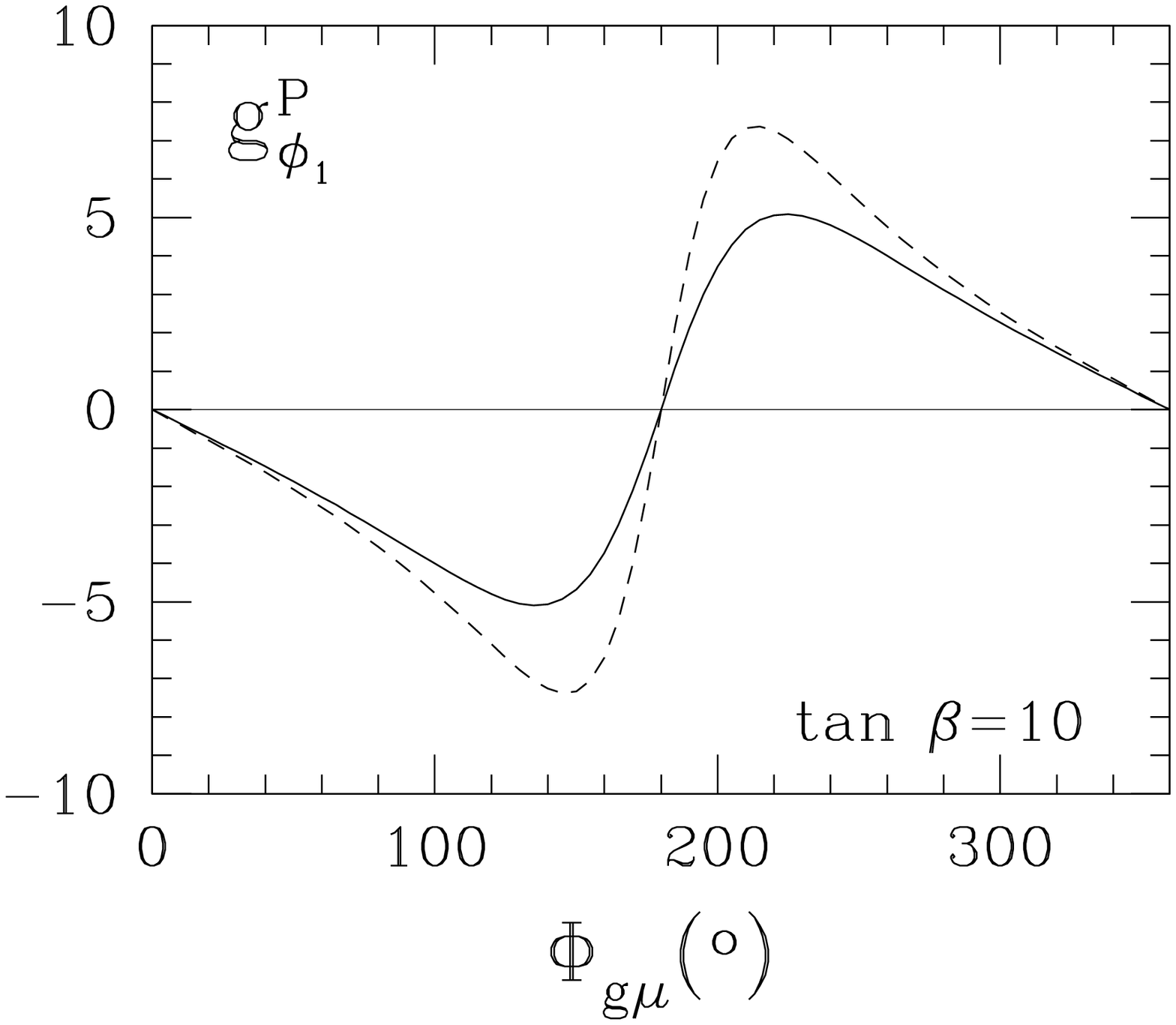}
\end{center}
\vspace{-0.4cm}
\begin{center}
\includegraphics[width=7.3cm]{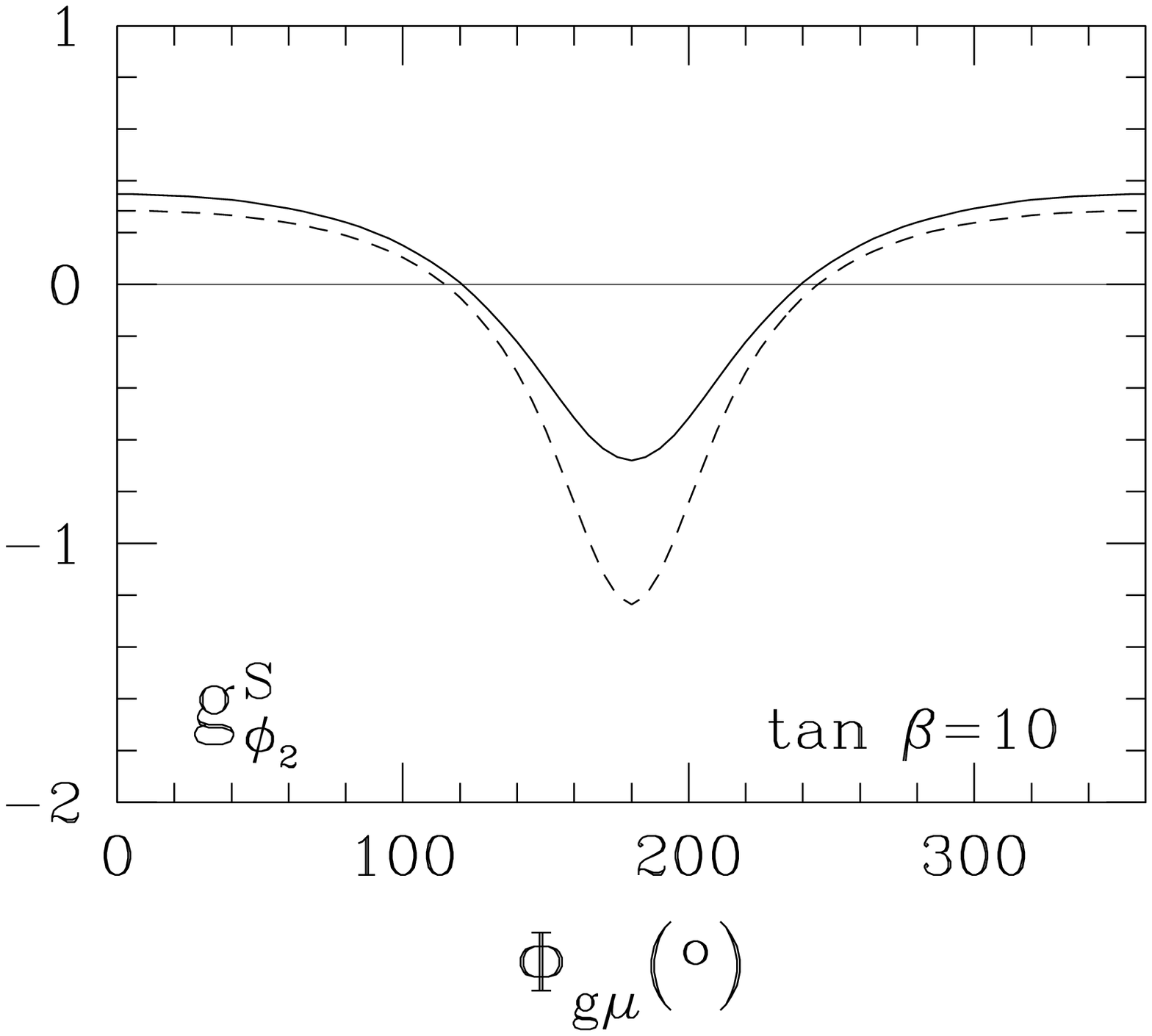}
\hspace{5mm}
\includegraphics[width=7.3cm]{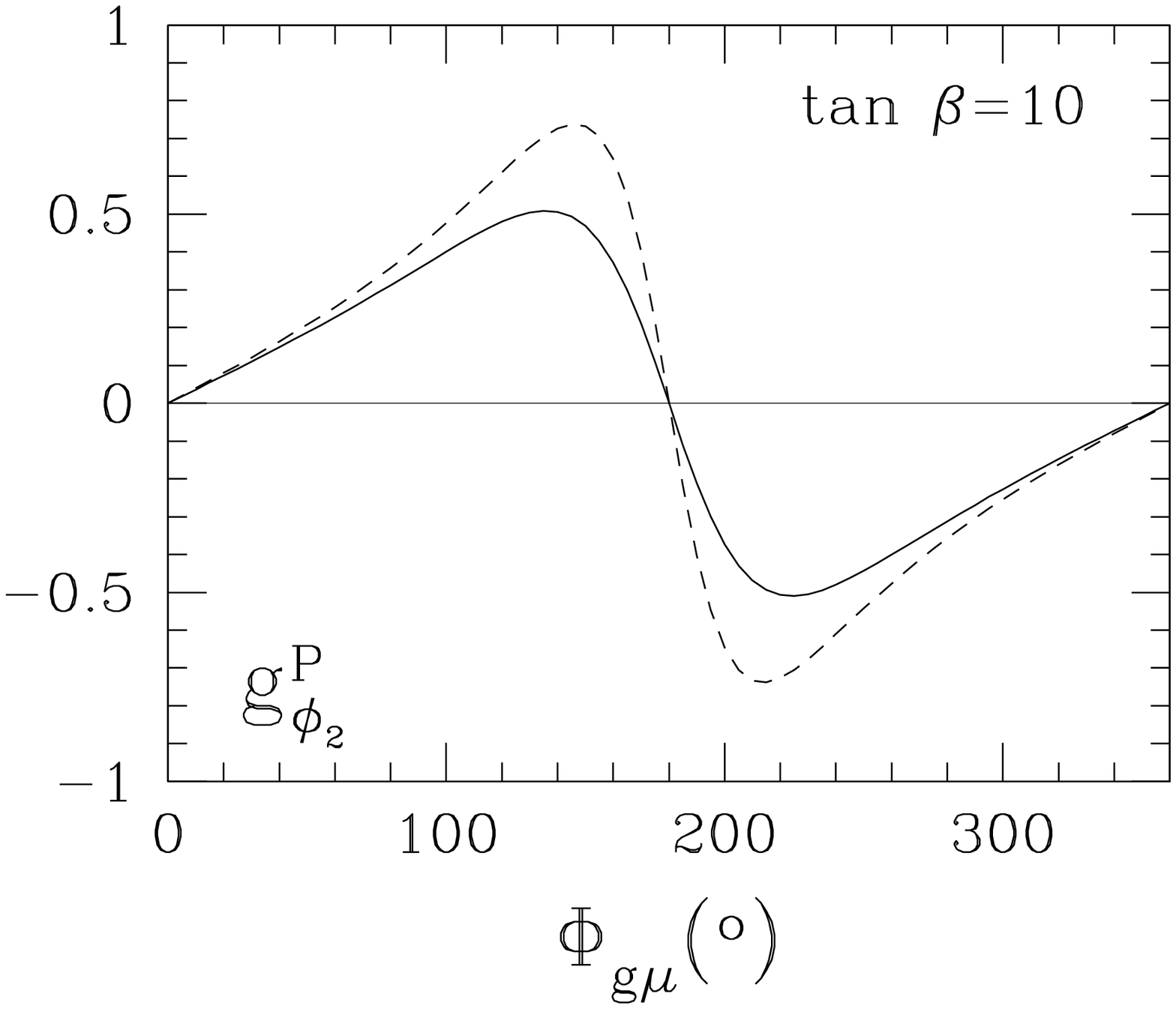}
\end{center}
\vspace{-0.4cm}
\begin{center}
\includegraphics[width=7.3cm]{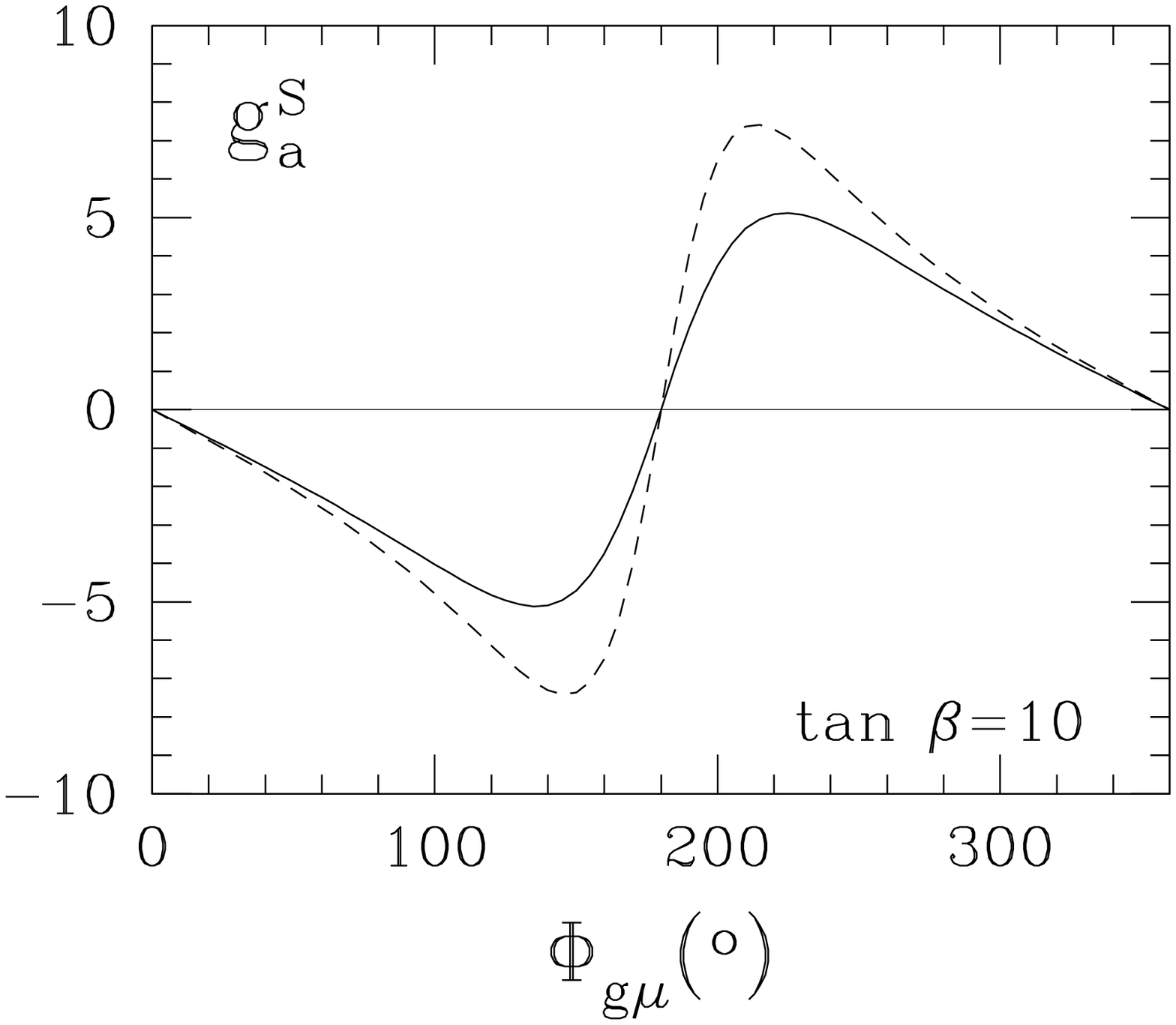}
\hspace{5mm}
\includegraphics[width=7.3cm]{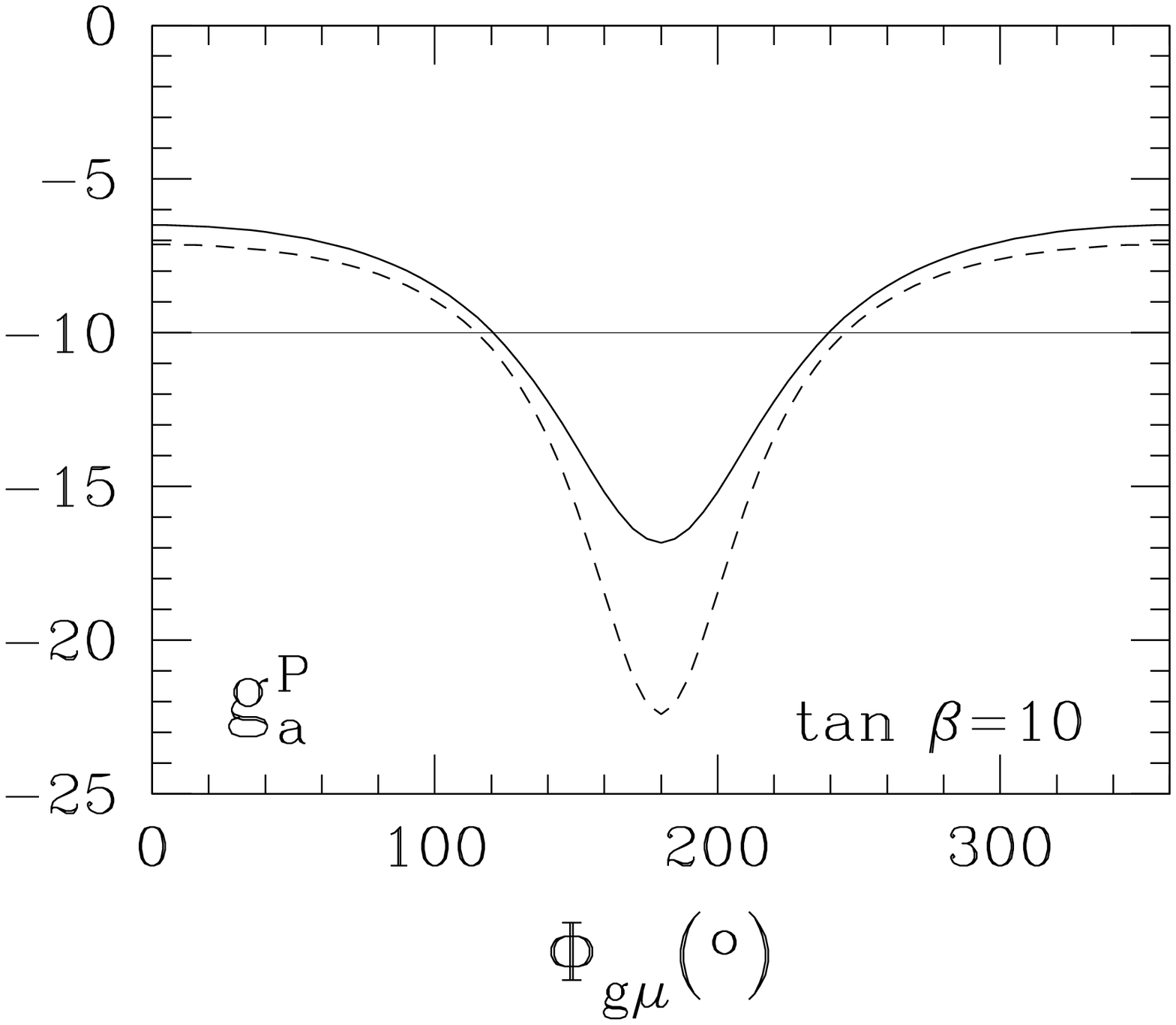}
\end{center}
\vspace{-0.4cm}
\caption{{\small \it Couplings $g_{\phi}^{S,P}$ vs. $\Phi_{g\mu}$, for
 the spectrum in Eq.~(\ref{eq:CPXpara}) with $c_A=c_\mu=1$, 
 $M_{\rm SUSY}=0.5\,$TeV, and $\tan\beta=10$. The solid lines are for
 $\Phi_{A\mu}=0^{\rm o}$, the dashed ones for
 $\Phi_{A\mu}=180^{\rm o}$.  The horizontal lines indicate the values
 of the uncorrected couplings.}}
\label{fig:ghbb}
\end{figure}

We turn now to consider the $b$-$b$-$H_i$ couplings.  Once the
threshold corrections to the $b$-quark mass are included, the
effective Lagrangian for the interaction of the neutral Higgs boson to
$b$ quarks can be written as
\begin{equation}
{\cal L}  \ =\ 
-\frac{m_b}{v}\, \bar{b}
   \left(g^S_\phi+ig^P_\phi\gamma_5\right) b \,\phi \,,
\label{eq:effLag}
\end{equation}
with $\phi=(\phi_1,\phi_2, a)$, where $\phi_1$ and $\phi_2$ are the 
CP-even parts of the neutral components of the two doublets:
\begin{equation}
  H_d^0 = \frac{1}{\sqrt{2}}\left(v_d +\phi_1 +i a_1 \right)\,, 
\quad \quad \quad
  H_u^0 = \frac{1}{\sqrt{2}}\left(v_u +\phi_2 +i a_2 \right)\,,
\label{eq:Hcomposition}
\end{equation}
and $a$ is a combination of the two CP-odd components $a_1$ and $a_2$,
$a =- a_1 \sin \beta + a_2 \cos \beta$. The orthogonal combination,
$G^0=a_1\cos\beta+a_2\sin\beta$, is the Goldstone mode.  The couplings
$g_\phi^{S,P}$ are~\cite{EDMandCOUPLINGS} 
\begin{equation}
\begin{array}{ll}
g_{\phi_1}^S \  =  \ \displaystyle{\frac{1}{\cos\beta}\,
                     \real{\left(\frac{1}{R_b}\right)}}\,,               
&\quad \quad 
g_{\phi_1}^P \  =  \ \displaystyle{\frac{\tan\beta}{\cos\beta}\,
                     \imag{\left(\frac{\kappa_b}{R_b}\right)}}\,,      
\nonumber \\[1.1ex]
g_{\phi_2}^S \  =  \ \displaystyle{\frac{1}{\cos\beta}\,
                     \real{\left(\frac{\kappa_b}{R_b}\right)}}\,,        
&\quad \quad 
g_{\phi_2}^P \  =  \ \displaystyle{-\frac{1}{\cos\beta}\,
                     \imag{\left(\frac{\kappa_b}{R_b}\right)}}\,,
\nonumber \\[1.1ex]
g_a^S        \  =  \ \displaystyle{(\tan^2\beta+1)\,
                     \imag{\left(\frac{\kappa_b }{R_b}\right)}}\,,       
&\quad \quad 
g_a^P        \  =  \ \displaystyle{-\real{\left(
                     \frac{\tan\beta-\kappa_b}{R_b}\right)}}\,,
\end{array}
\label{eq:g-phi-sp}
\end{equation}
and for values of $\tan \beta$ such that 
$\vert \kappa_b \vert \tan \beta \sim 1$ (with one of the 
two, or both possibilities: $\real{(\kappa_b)} \tan \beta \sim 1$, 
$\imag{(\kappa_b)} \tan \beta \sim 1$) reduce to 
\begin{equation}
\begin{array}{ll}
g_{\phi_1}^S \  =  \ \displaystyle{
   \frac{\tan \beta}{\vert R_b \vert^2}
   \left[{1+\real{\left(\kappa_b\right)}\tan\beta}\right]}
\,,
&\quad \quad 
g_{\phi_1}^P \  =  \ \displaystyle{
   \frac{\tan \beta}{\vert R_b \vert^2}\, 
   \left[{\imag{\left(\kappa_b\right)}\tan \beta }\right]}
\,,             \nonumber \\[2.5ex]
g_{\phi_2}^S \  =  \ \displaystyle{
   \frac{1}{\vert R_b \vert^2}
  \left[{\real{\left(\kappa_b\right)}\tan \beta +
      \vert \kappa_b \vert^2 \tan^2 \beta} \right]}
\,,      
&\quad \quad 
g_{\phi_2}^P \  =  \ \displaystyle{
 - \frac{1}{\vert R_b \vert^2} \,
   \left[{\imag{\left(\kappa_b\right)}}\tan \beta \right]}
\,,             \nonumber\\[2.5ex]     
g_a^S        \  =  \ \displaystyle{
   \frac{\tan \beta}{\vert R_b \vert^2} \,
   \left[{\imag{\left(\kappa_b\right)}}\tan \beta \right]}
\,,      
&\quad \quad 
g_a^P        \  =  \ \displaystyle{
   -\frac{\tan \beta}{\vert R_b \vert^2}
 \left[1 +\real{\left(\kappa_b \right)}\tan \beta\right]}
\,.            
\end{array}
\label{eq:g-phi-sp2}
\end{equation}
Before proceeding further, we list in the following some of the
interesting features of these couplings. We illustrate them
numerically for a specific CPX spectrum, i.e.  with $c_A=c_\mu=1$,
$M_{\rm SUSY}=0.5\,$TeV, and $\tan\beta =10$, which is a value not
plagued by the problem of a tachyonic $\tilde{b}_1$ squark discussed
earlier. For visual clarity we also illustrate some of these
features in Fig.~\ref{fig:ghbb}.
\\[1.5ex]
\noindent
$\bullet$ 
If no threshold corrections are included, the only nonvanishing
couplings are $g_{\phi_1}^S=1/\cos\beta$ and $g_a^P=-\tan\beta$.  The
inclusion of these corrections affects these two couplings mainly
through the factor $\real{(1/R_b)}$, which is a suppression or an
enhancement factor, depending on the value of $\Phi_{g\mu}$, and
varies between~$\sim 1/(1+|\epsilon_g|\tan\beta)$ 
and~$\sim 1/(1-|\epsilon_g|\tan\beta)$, obtained for
$\Phi_{g\mu}=0^{\rm o}$ and $\Phi_{g\mu}=180^{\rm o}$, respectively.
We have used here the fact that $|\epsilon_g| \gg |\epsilon_H|$,
which, as previously discussed, is rather generic in our
scenarios. Notice also that the factor $\real{(1/R_b)}$ is larger
than~1 for
$\cos\Phi_{g\mu} \stackrel{<}{{}_\sim} - |\epsilon_g|\tan\beta$. For 
$\tan \beta=10$ and the spectrum specified above, this happens when 
$135^{\rm o} \lsim \Phi_{g\mu} \lsim 225^{\rm o}$, and 
$\vert g_{\phi_1}^S\vert $ and $\vert g_a^P \vert$ reach the
maximum of $15$ or $\sim 20$, depending on the value of $\Phi_{A\mu}$ 
(i.e. $0^{\rm o}$ or $180^{\rm o}$), at $\Phi_{g\mu}=180^{\rm o}$.
\\[1.5ex]
\noindent
$\bullet$ 
The inclusion of threshold corrections is responsible for the 
appearance of the other four couplings: $g_{\phi_2}^S$ and 
$g_{\phi_2}^P$, the smallish ones, {\it i.e.} without the overall 
factor of $\tan \beta$ that $g_{\phi_1}^S$, $g_a^P$ have (when  
$\vert \kappa_b \vert \tan \beta \sim 1$), $g_{\phi_1}^P$ and
$g_a^S$, the large ones, with the $\tan \beta$ factor.
\\[1.5ex]
\noindent
$\bullet$ 
If no CP phases are present, only $g_{\phi_1}^S$, $g_{\phi_2}^S$, 
and $g_a^P$ are nonvanishing and, again, the effect of the
threshold corrections is more significant for
$\Phi_{g\mu}=180^{\rm o}$ than for $\Phi_{g\mu}=0^{\rm o}$. 
\\[1.5ex]
\noindent
$\bullet$ 
As the phase $\Phi_{g\mu}$ varies, the moduli of the couplings 
$g_{\phi_1}^P$, $g_{\phi_2}^P$, and $g_a^S$ reach their extremal 
values at 
$\cos\Phi_{g\mu}\approx
 {-2|\epsilon_g|\tan\beta}/{(1+|\epsilon_g|^2\tan^2\beta)}$. In the 
specific case considered here, this is~5 or~8 for $|g_{\phi_1}^P|$ and
$|g_a^S|$ at $\Phi_{g\mu}\approx 150^{\rm o}$ and~$210^{\rm o}$,
depending on the value of $\Phi_{A\mu}$. In contrast,
$|g_{\phi_2}^P|$, as well as $|g_{\phi_2}^S|$, never exceed~$1.2$.

The states $\phi_1$, $\phi_2$, and $a$ are not yet the neutral Higgs
boson mass eigenstates. Their real and symmetric 3$\times$3 matrix
${\cal M}^2_H$ has nonvanishing entries that mix the two states
$\phi_1$ and $\phi_2$ as well as nonvanishing CP-violating entries
that mix $a$ with $\phi_1$ and $\phi_2$ and are proportional to
$\Phi_{A\mu}$. The diagonalization of this matrix through an
orthogonal matrix $O$,
\begin{eqnarray}
 O^T{\cal M}^2_H\, O  =
             {\sf diag}\,(m^2_{H_1},m^2_{H_2},m^2_{H_3})\,,
\label{eq:Omix}
\end{eqnarray}
yields the three eigenstates ${H_1},{H_2},{H_3}$, ordered for
increasing value of their masses.  In the limit of vanishing
$\Phi_{A\mu}$, $H_1$ is $h$, $H_2$ and $H_3$ are $H$ and $A$, or vice
versa, $A$ and $H$, depending on the values of $m_A$ and $m_H$. After
this rotation, the effective Lagrangian of Eq.~(\ref{eq:effLag})
becomes
\begin{equation}
{\cal L} \ \to \ -\frac{m_b}{v}\, \bar{b}
  \left(g^S_{H_i}+ig^P_{H_i}\gamma_5\right) b \,H_i \,,
\label{eq:effLagrotated}
\end{equation}
where $g_{H_i}^{S}$ and $g_{H_i}^{P}$ are
\begin{equation}
 g_{H_i}^{S}=O_{\alpha i}\,g_{\alpha}^{S}\,,
\quad \quad
 g_{H_i}^{P}=O_{\alpha i}\,g_{\alpha}^{P}\,,
\label{eq:g-hi}
\end{equation}
with the index $\alpha$ running over $(\phi_1,\phi_2,a)$ and $i$ over
$(1,2,3)$.

We are now in a position to evaluate the production cross sections of
the neutral Higgs bosons $H_i$ via $b$-quark fusion at hadron
colliders. They can be expressed as:
\begin{equation}
\sigma({\rm had}_1 {\rm had}_2\rightarrow b\bar{b}\rightarrow H_i) =
 {\sigma}(b\bar{b}\rightarrow H_i)
 \int_{\tau_i}^1 {\rm d}x
 \left[\frac{\tau_i}{x}\, 
        b_{{\rm had}_1}\!(x,Q) \,
        \bar{b}_{{\rm had}_2}\!\!\left(\frac{\tau_i}{x},Q\!\right)
        +(b\leftrightarrow \bar{b}) \right]  \,,
\label{eq:Xsection}
\end{equation}
where $b_{{\rm had}_i}(x,Q)$ and $\bar{b}_{{\rm had}_i}(x,Q)$ are the
$b$- and $\bar{b}$-quark distribution functions in the hadron 
${\rm had}_i$, $Q$ is the factorization scale, and $\tau_i$ the
Drell--Yan variable $\tau_i=m_{H_i}^2/s$, with $s$ the invariant
hadron-collider energy squared. Finally, the partonic cross section is
\begin{equation}
{\sigma}(b\bar{b}\rightarrow H_i)=
         \frac{m_b^2}{v^2}\frac{\pi}{6m_{H_i}^2}
\left[(g_{H_i}^S)^2+ (g_{H_i}^P)^2\,\right] \,.
\label{eq:partonicXsect}
\end{equation}

The sum of the couplings squared 
\begin{equation}
\left[(g_{H_i}^S)^2 + (g_{H_i}^P)^2 \,
\right] = 
\left[\left(O_{a i} g_{a}^{S} +\!
            O_{\phi_1 i} g_{\phi_1}^{S} +\!
            O_{\phi_2 i} g_{\phi_2}^{S}
      \right)^2  +    
      \left(O_{a i} g_{a}^{P} +\!
            O_{\phi_1 i} g_{\phi_1}^{P} +\!
            O_{\phi_2 i} g_{\phi_2}^{P}
      \right)^2  
\right]
\label{eq:couplingsum}
\end{equation}
reduces, in the limit in which no threshold corrections are included, 
to 
\begin{equation}
\left[(g_{H_i}^S)^2 + (g_{H_i}^P)^2 \right]\vert_{no-thresh-corr} = 
\left[ O_{\phi_1 i}^2  (g_{\phi_1}^{S})^2 +
       O_{a i}^2       (g_{a}^{P})^2         
\right]\,,
\label{eq:nothreshsum}
\end{equation}
with $g_{\phi_1}^S=1/\cos\beta$ and $g_a^P=-\tan\beta$.  For $i=1$, in 
the limit of large $m_{H^\pm}$, $O_{\phi_1 1} \to \cos\beta$ and 
$O_{a 1} \to 0$, leading to the usual Standard Model coupling of the
lightest neutral Higgs boson to $b$ quarks, in both, CP-conserving and
CP-violating scenarios.  That is, the overall $\tan \beta$ dependence
of the coupling $g_{\phi_1}^S$ is killed by the mixing element
$O_{\phi_1 1}$.

Through the inclusion of the threshold corrections to the $b$ quark in
CP-conserving scenarios, the element $g_{\phi_2}^S$ gets switched
on. Being one of the small couplings, i.e. one without an overall
$\tan \beta$ factor when $\tan \beta$ is large, $g_{\phi_2}^S$ is not
expected to produce great numerical deviations in the values of the
cross sections at large $\tan \beta$.  Deviations of $O(1)$ can only
be induced by the factors $1/\vert R_b \vert^4$, which enhance or
suppress the uncorrected cross sections, depending on the sign of the
threshold corrections.  It is however in CP-violating scenarios, when
two couplings with an overall $\tan \beta$ dependence are switched on,
$g_{\phi_1}^P$ and $g_a^S$, that the pattern of Higgs production
through $b$-quark fusion can become very different.  This, of course,
if the projection of Higgs boson current-eigenstates to
mass-eigenstates is not particularly destructive.

To analyze the behaviour of the sum in Eq.~(\ref{eq:couplingsum}), we 
use the approximation: 
\begin{equation}
\left[(g_{H_i}^S)^2 +(g_{H_i}^P)^2 \right] \, \approx\,
\left[O_{\phi_1 i}^2 + O_{a i}^2   \right]
\left[(g_a^S)^2 +(g_a^P)^2         \right]    
\,.
\label{eq:approximation1}
\end{equation}
In this, we have parametrized the sum $[(g_{H_i}^S)^2 +(g_{H_i}^P)^2]$
in terms of the only elements $O_{\phi_1 i}$ and $O_{a i}$ that appear
in Eq.~(\ref{eq:nothreshsum}), i.e. in the case in which no threshold
corrections to the $b$-quark mass are included.  This approximation
relies on the properties of the couplings $g_{\phi}^{S,P}$ discussed
above, i.e. on the fact that $g_{\phi_1}^S\approx -g_a^P$,
$g_{\phi_1}^P\approx g_a^S$, and that $g_{\phi_2}^{S}$ and
$g_{\phi_2}^{P}$ can be neglected. In the scenarios considered here,
this approximation turns out to be valid for most of the relevant
values of the phases $\Phi_{g\mu}$ and $\Phi_{A\mu}$.

\begin{figure}[t]
\begin{center}
\includegraphics[width=7.3cm]{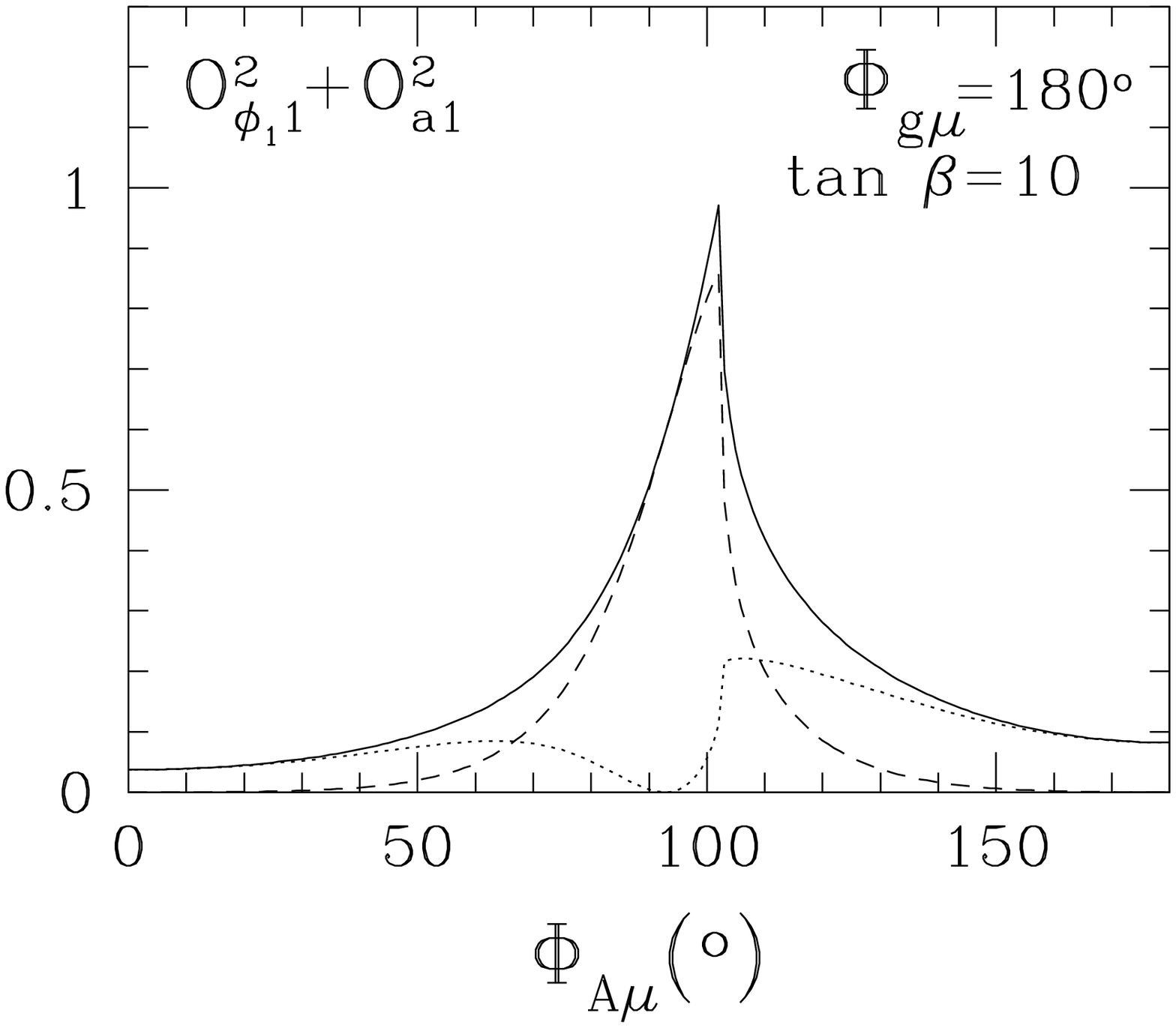}
\end{center}
\vspace{-0.4cm}
\begin{center}
\includegraphics[width=7.3cm]{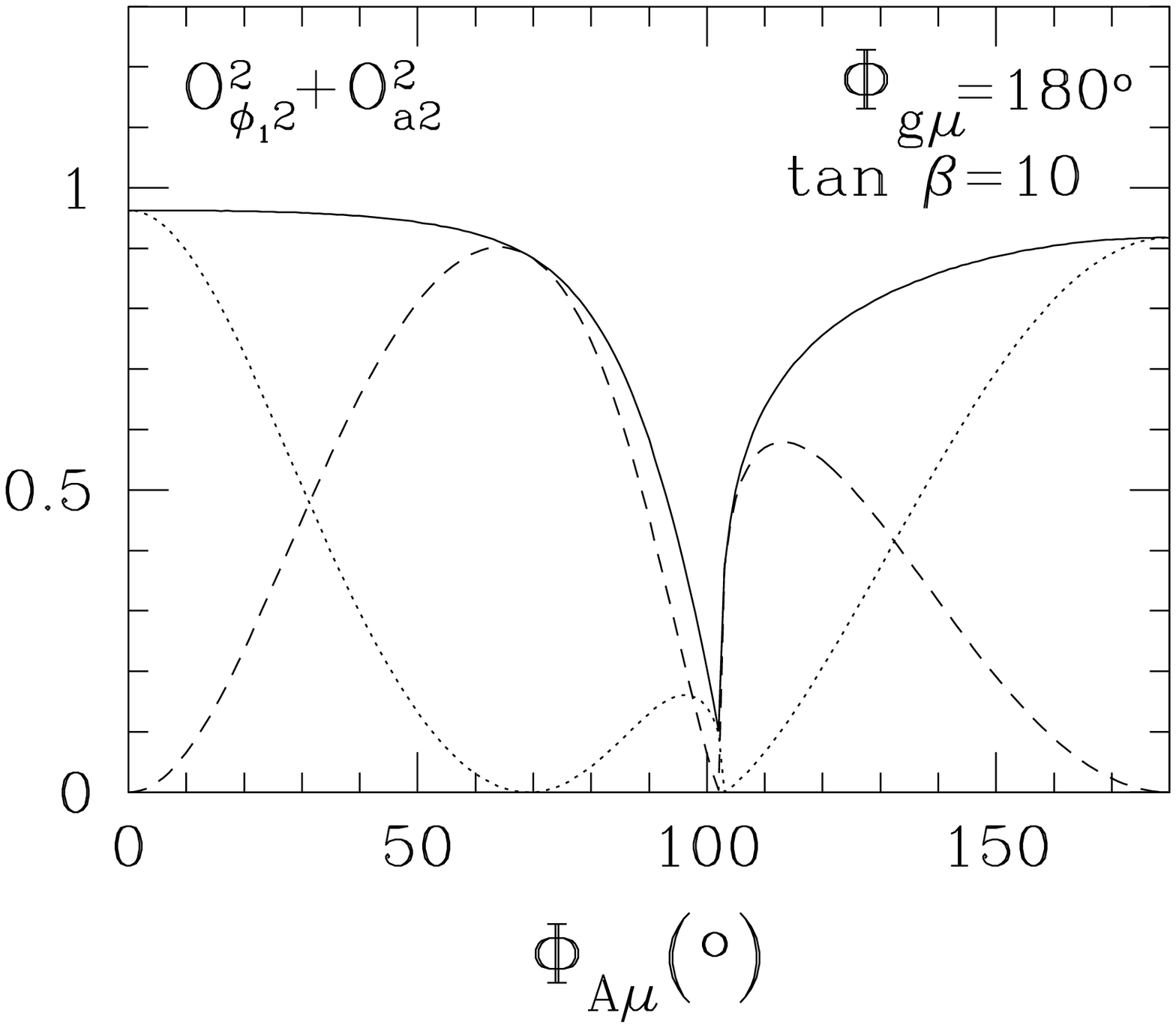}
\hspace{5mm}
\includegraphics[width=7.3cm]{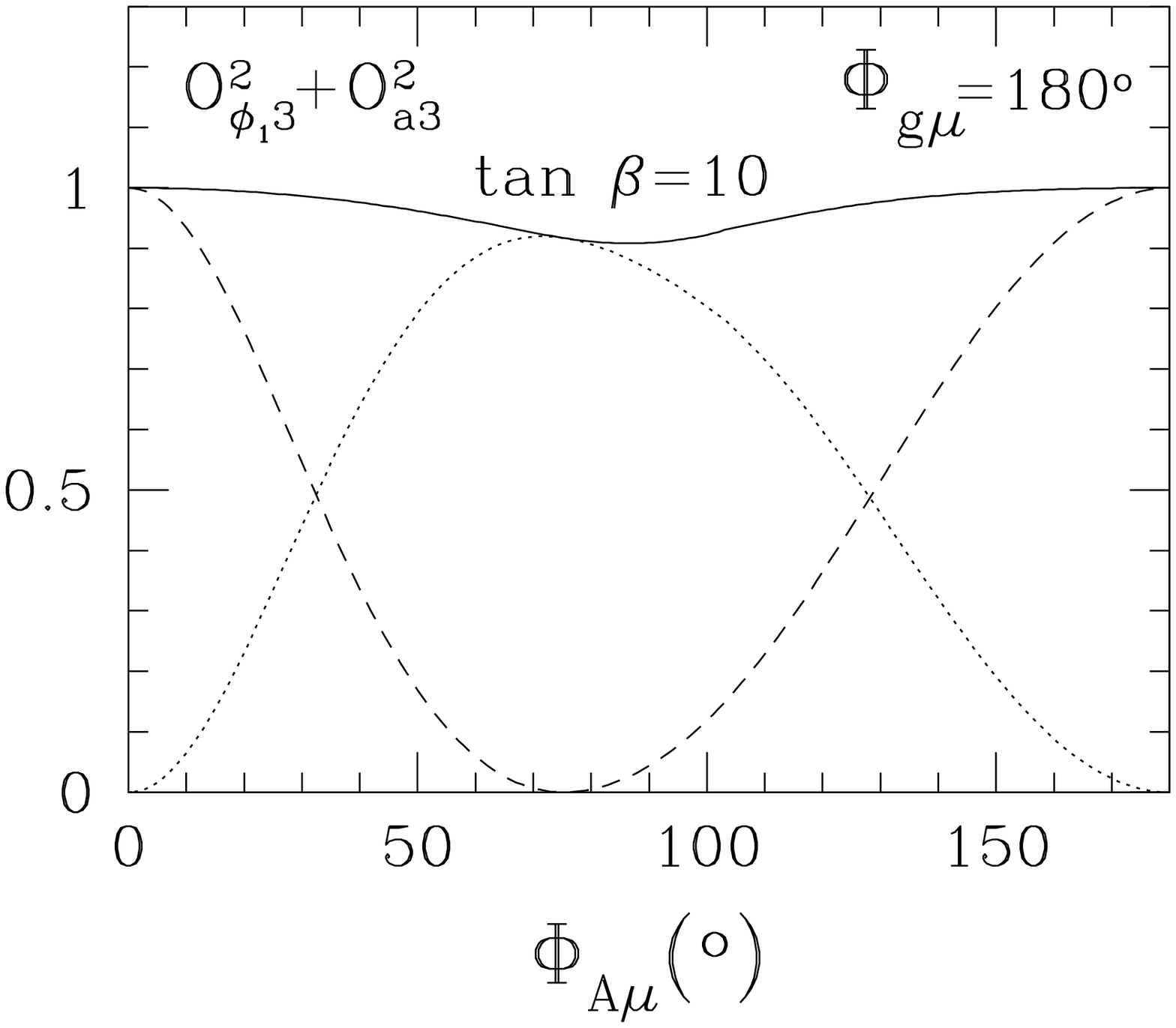}
\end{center}
\vspace{-0.4cm}
\caption{{\small \it The sums  $O_{\phi_1i}^2+O_{ai}^2$ vs.  $\Phi_{A\mu}$, 
 for the spectrum of Eq.~(\ref{eq:CPXpara}) with $c_A=c_\mu=1$,
 $\tan\beta=10$,  as in Fig.~\ref{fig:ghbb}, and  $m_{H_1}=115\,$GeV, 
 $\Phi_{g\mu}=180^{\rm o}$. The dashed lines show $O_{ai}^2$; the dotted
 ones, $O_{\phi_1i}^2$.}}
\label{fig:Omix}
\end{figure}

To strengthen our point, we show explicitly the mixing elements
$O_{\phi i}$ for the scenarios already considered in
Fig.~\ref{fig:ghbb}. We show them in Fig.~\ref{fig:Omix} versus
$\Phi_{A\mu}$, but with $\Phi_{g\mu}$ fixed at $180^{\rm o}$. As
already remarked, the elements $O_{\phi i}$ have a rather weak
dependence on $\Phi_{g\mu}$, coming from the two-loop corrections to
the Higgs potential.  The charged Higgs boson mass is solved in order
to have $m_{H_1}=115\,$GeV, for all values of $\Phi_{A\mu}$ and
$\Phi_{g\mu}$, and it is therefore different for different values of
these phases. As a consequence, also $m_{H_2}$ and $m_{H_3}$ are
varying quantities. Notice that, for 
$\Phi_{A \mu}\approx 100^{\rm o}$, $H_1$ is predominantly the CP-odd
$a$ boson, whereas $H_2$ and $H_3$ are mainly $\phi_2$ and $\phi_1$,
respectively. For these values of $\Phi_{A \mu}$, the charged Higgs
boson is relatively light, i.e. it has a mass $\lsim 200\,$GeV.

A comparison of Figs.~\ref{fig:ghbb} and~\ref{fig:Omix} indicates that
the approximation in Eq.~(\ref{eq:approximation1}) is probably not
adequate when $[O_{\phi_1 i}^2 + O_{a i}^2]\ll 1$, since the neglect
of the elements $g_{\phi_2}^{S}$ and $g_{\phi_2}^{P}$ cannot be
justified. Therefore, we keep the exact expression in
Eq.~(\ref{eq:couplingsum}) for the calculation of the production cross
sections, but we use this approximation in order to clarify the
$\tan \beta$ dependence of the sum $[(g_{H_i}^S)^2 + (g_{H_i}^P)^2]$. 
By substituting in Eq.~(\ref{eq:approximation1}) the expression of the 
couplings $g_{\phi}^{S,P}$ at large $\tan \beta$, we obtain:
\begin{equation}
\left[(g_{H_i}^S)^2 + (g_{H_i}^P)^2 \right] 
\, \approx\,
\left(O_{\phi_1 i}^2 +  O_{a i}^2 \right) 
\frac{\tan^2\beta}{|R_b|^2}  \,.
\label{eq:approximation2}
\end{equation}
That is to say, the terms in $\tan^4 \beta$ cancel identically. It
remains only a factor of $\tan^2 \beta$ suppressed or enhanced by
$1/|R_b|^2$, in which all dependence on $\Phi_{g \mu}$ gets
concentrated. Whether this factor of $\tan^2 \beta$ is inherited
unharmed by the cross sections for all Higgs bosons $H_i$ depends on
the matrix elements $O_{a i}$ and $O_{\phi_1 i}$. For the values of
$\Phi_{A \mu}$ for which the charged Higgs boson is not very heavy,
and the three Higgs boson states are significantly mixed, we expect
this to be the case, in general for all three states, and in
particularly also for the lightest one. That is, for 
$\Phi_{A \mu}\approx 100^{\rm o}$, we expect a large enhancement of
the production cross section for ${H_1}$, but presumably a suppression
of that for ${H_2}$, as an inspection of Fig.~\ref{fig:Omix} suggests.

\begin{figure}[t]
\begin{center}
\includegraphics[width=7.3cm]{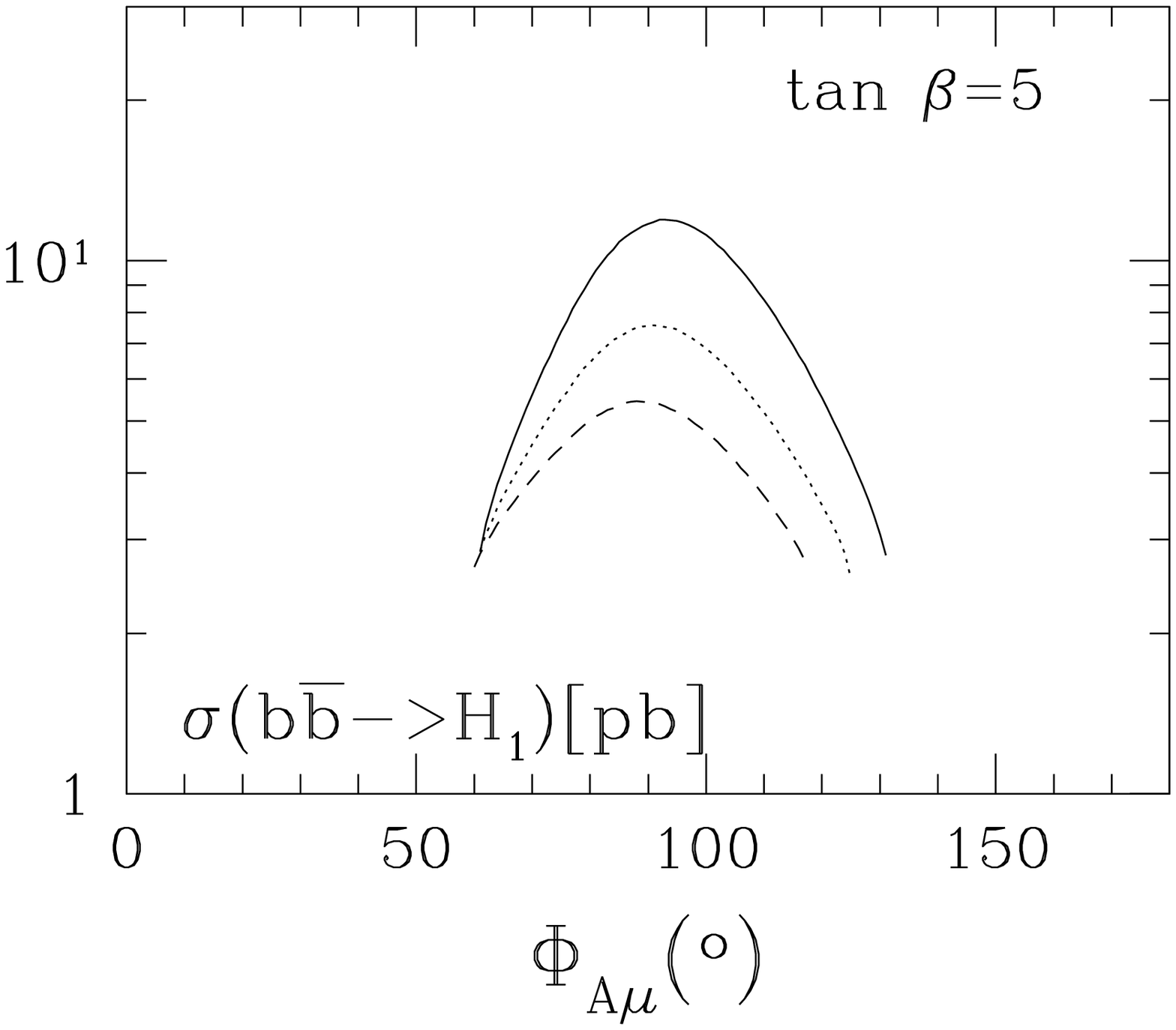}
\hspace{5mm}
\includegraphics[width=7.3cm]{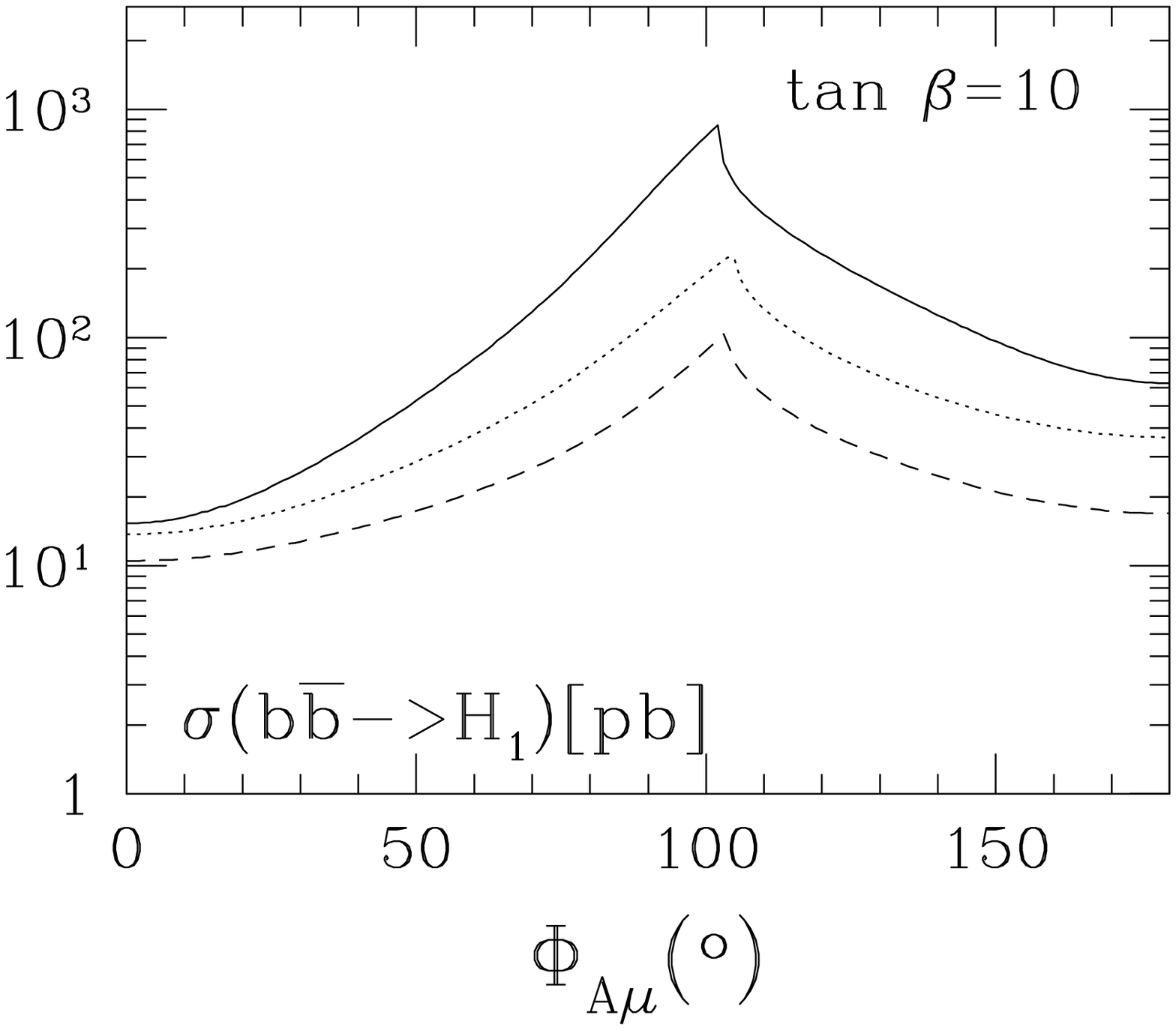}
\end{center}
\vspace{-0.4cm}
\begin{center}
\includegraphics[width=7.3cm]{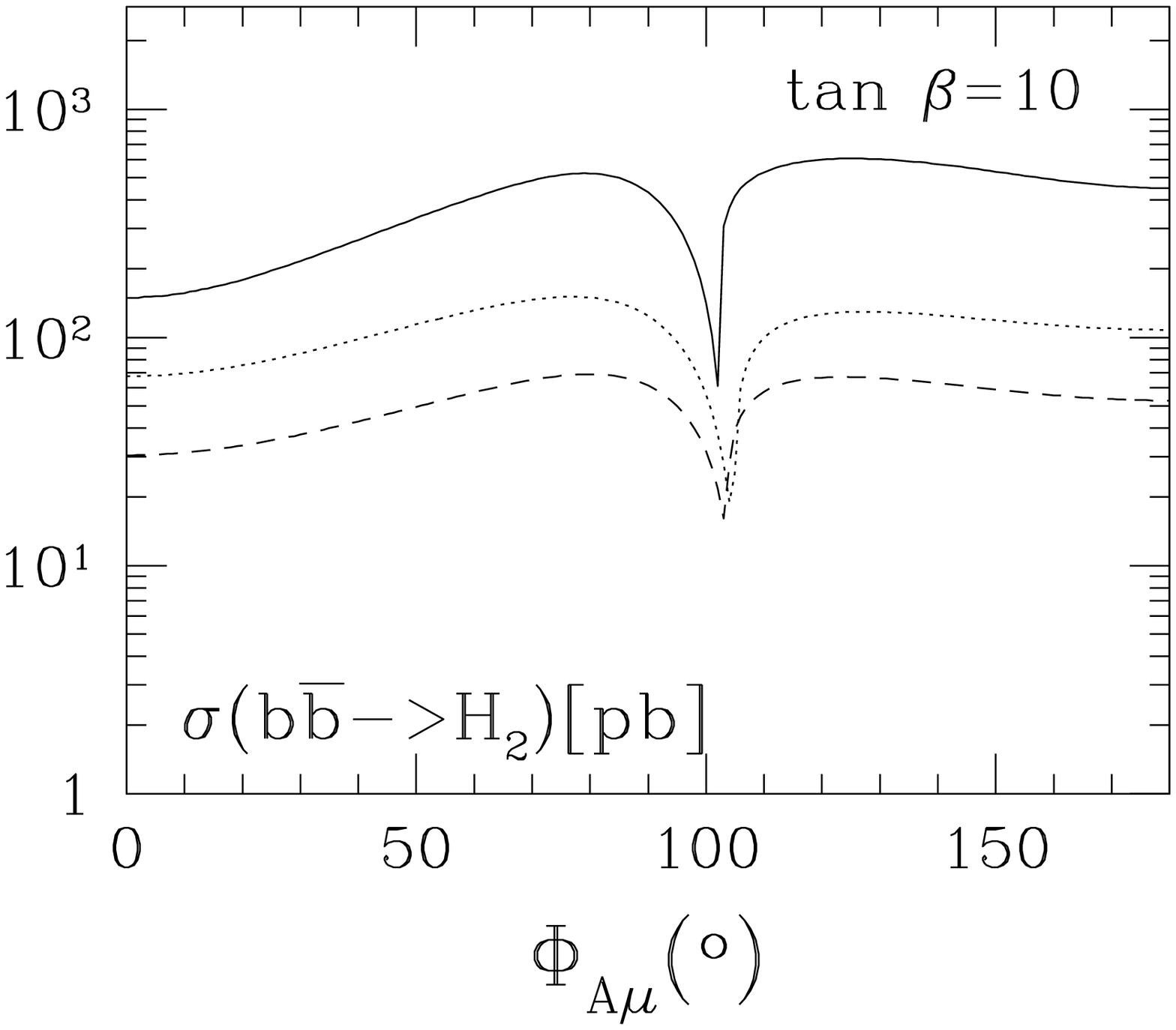}
\hspace{5mm}
\includegraphics[width=7.3cm]{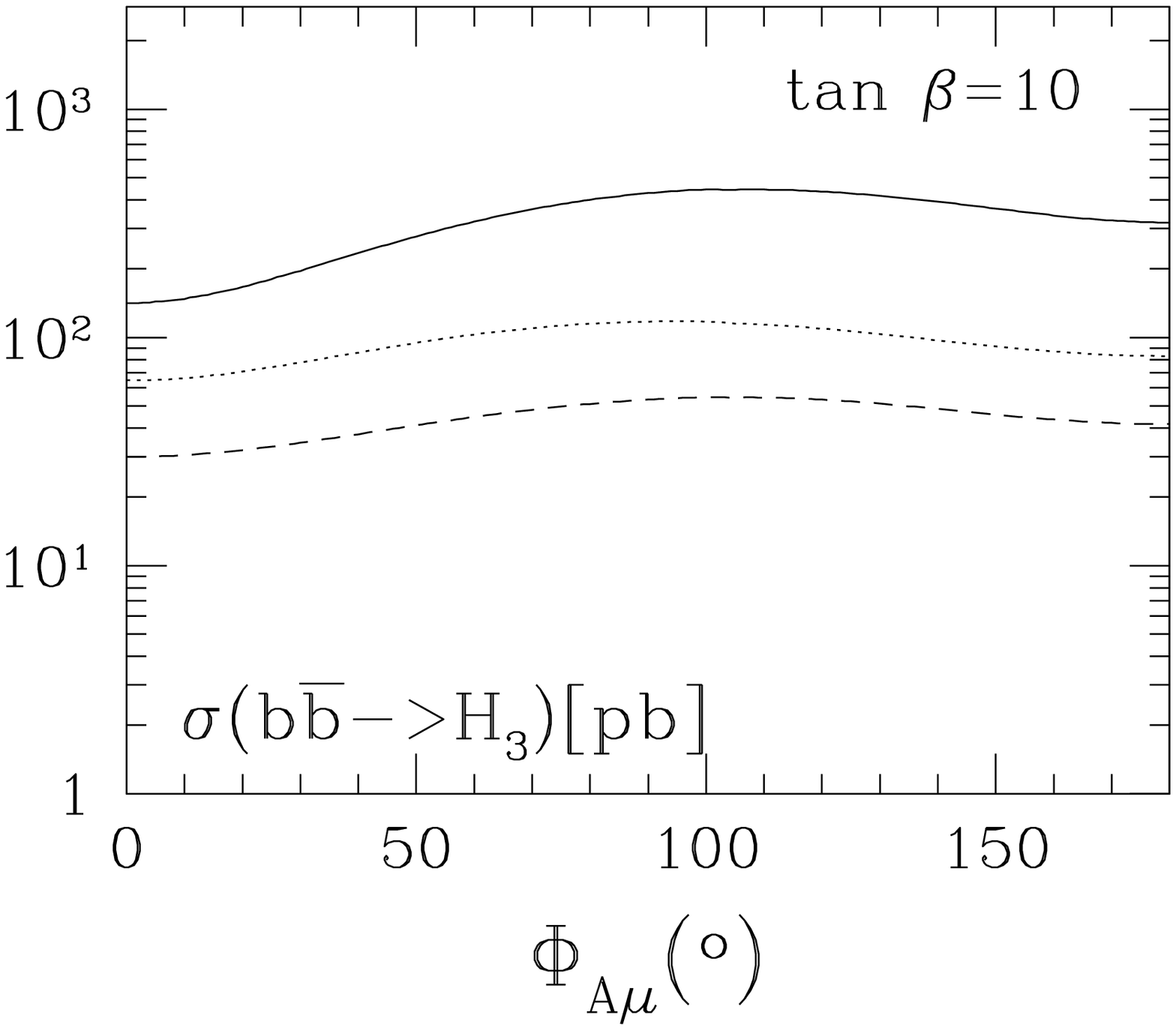}
\end{center}
\vspace{-0.4cm}
\caption{{\small \it Partonic cross sections for the $b$-quark fusion
 production of $H_1$, $H_2$, and $H_3$ vs. $\Phi_{A\mu}$, for
 $\Phi_{g\mu}=0^{\rm o}$~(dashed lines) and $180^{\rm o}$~(solid lines).
 The supersymmetric spectrum is that used also for Fig.~\ref{fig:ghbb}.
 In dotted lines are also the cross sections with no $m_b$-corrections.}}
\label{fig:partXsection}
\end{figure}

We show the partonic cross sections in Fig.~\ref{fig:partXsection}, 
as functions of $\Phi_{A\mu}$, for $\Phi_{g\mu} =0^{\rm o}$
(dashed lines) and $180^{\rm o}$ (solid lines). The dotted lines
indicate the cross sections without threshold corrections to the
$b$-quark mass.  The cross sections for all neutral Higgs bosons are
shown for $\tan\beta=10$, that for $H_1$ also for $\tan\beta=5$. As
mentioned, the two upper frames show cross sections for constant
$m_{H_1}$, i.e.  $m_{H_1}=115\,$GeV, which cannot actually be realized
for all values of $\Phi_{A\mu}$ in the case of $\tan \beta=5$
($m_{H_1}$ tends to be lighter for $\Phi_{A\mu} \lsim 50^o$ and
$\Phi_{A\mu} \gsim 130^o$).  In contrast, in the two lower frames,
$m_{H_2}$ and $m_{H_3}$ are different for different values of
$\Phi_{A\mu}$: $m_{H_2}(m_{H_3})$ reaches the maximum 
of~$\sim 240(250)\,$GeV at $\Phi_{A\mu}=0^{\rm o}$ and
$\Phi_{g\mu}=180^{\rm o}$, the minimum of~$\sim 120 (150)\,$GeV
for~$\Phi_{A\mu} \sim 90^{\rm o}$, nearly independently
of~$\Phi_{g\mu}$.  All cross sections show the typical pattern already
observed for~$O_{\phi_1i}^2+O_{ai}^2$ in Fig.~\ref{fig:Omix}, with
modulations in the cases~$i=2,3$, due to varying values of the
corresponding Higgs masses.  Notice the increase of almost two orders
of magnitude in the cross section for $H_1$, at 
$\Phi_{A\mu}\approx 100^{\rm o}$, when the value of $\tan \beta$ is
only doubled. Indeed, for $i=1$, the sum $O_{\phi_1i}^2+O_{ai}^2$ is
larger for~$\tan \beta =10$ than for~$\tan \beta=5$, when larger
values of $m_{H^\pm}$ are needed to ensure that $m_{H_1}=115\,$GeV.
Moreover, at $\tan \beta =5$ the eigenstate $H_1$ has still a large
CP-even component.

Overall, the $m_b$ corrections increase the cross sections 
$\sigma(b \bar{b}\to H_i)$ with respect to the uncorrected ones for
$\Phi_{g\mu} = 180^{\rm o}$, decrease them for 
$\Phi_{g\mu} = 0^{\rm o}$.  The impact of these corrections, is in
general always large, producing enhancements up to one order of
magnitude and suppressions down to $-60\%$.  Notice that the
corrections remain surprisingly large for $H_2$ and $H_3$, also in
the CP-conserving cases
$\Phi_{g\mu} = 180^{\rm o}$, $\Phi_{A\mu} = 0^{\rm o}$  ($+120\%$); 
$\Phi_{g\mu} = 180^{\rm o}$, $\Phi_{A\mu} = 180^{\rm o}$ ($+300\%$); 
as well as 
$\Phi_{g\mu} = 0^{\rm o}$, $\Phi_{A\mu} = 0^{\rm o}, 180^{\rm o}$
($-50\%$).  
They are more modest for $H_1$ at 
$\Phi_{g\mu} = 0^{\rm o}, 180^{\rm o}$,
$\Phi_{A\mu} = 0^{\rm o}$ (i.e. $\sim \pm 20\%$), but still 
$+70\%$ and $-50\%$ at 
$\Phi_{g\mu} = 0^{\rm o}, 180^{\rm o}$ and 
$\Phi_{A\mu} = 180^{\rm o}$.

\begin{figure}[t]
\begin{center}
\includegraphics[width=7.3cm]{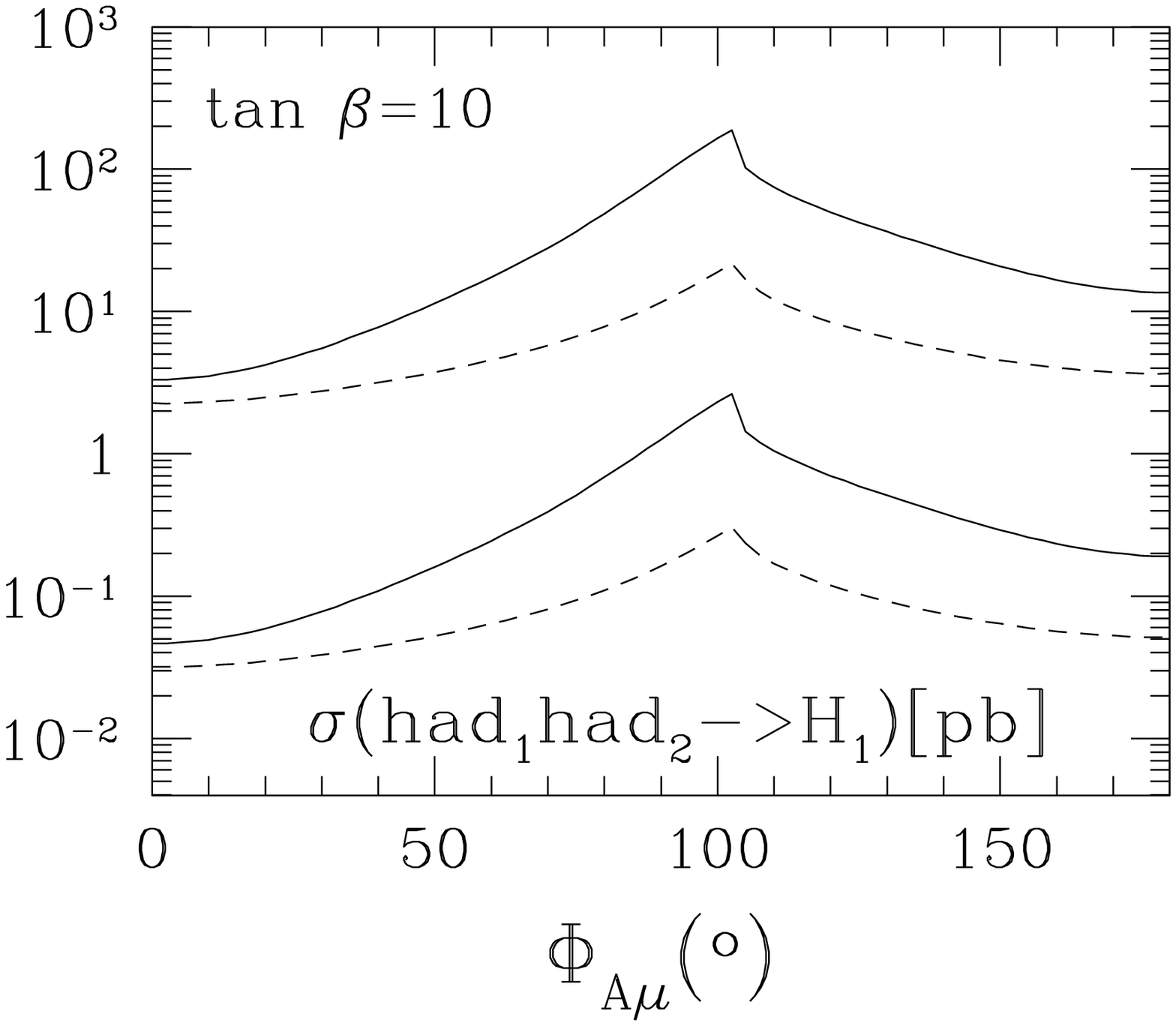}
\end{center}
\vspace{-0.4cm}
\begin{center}
\includegraphics[width=7.3cm]{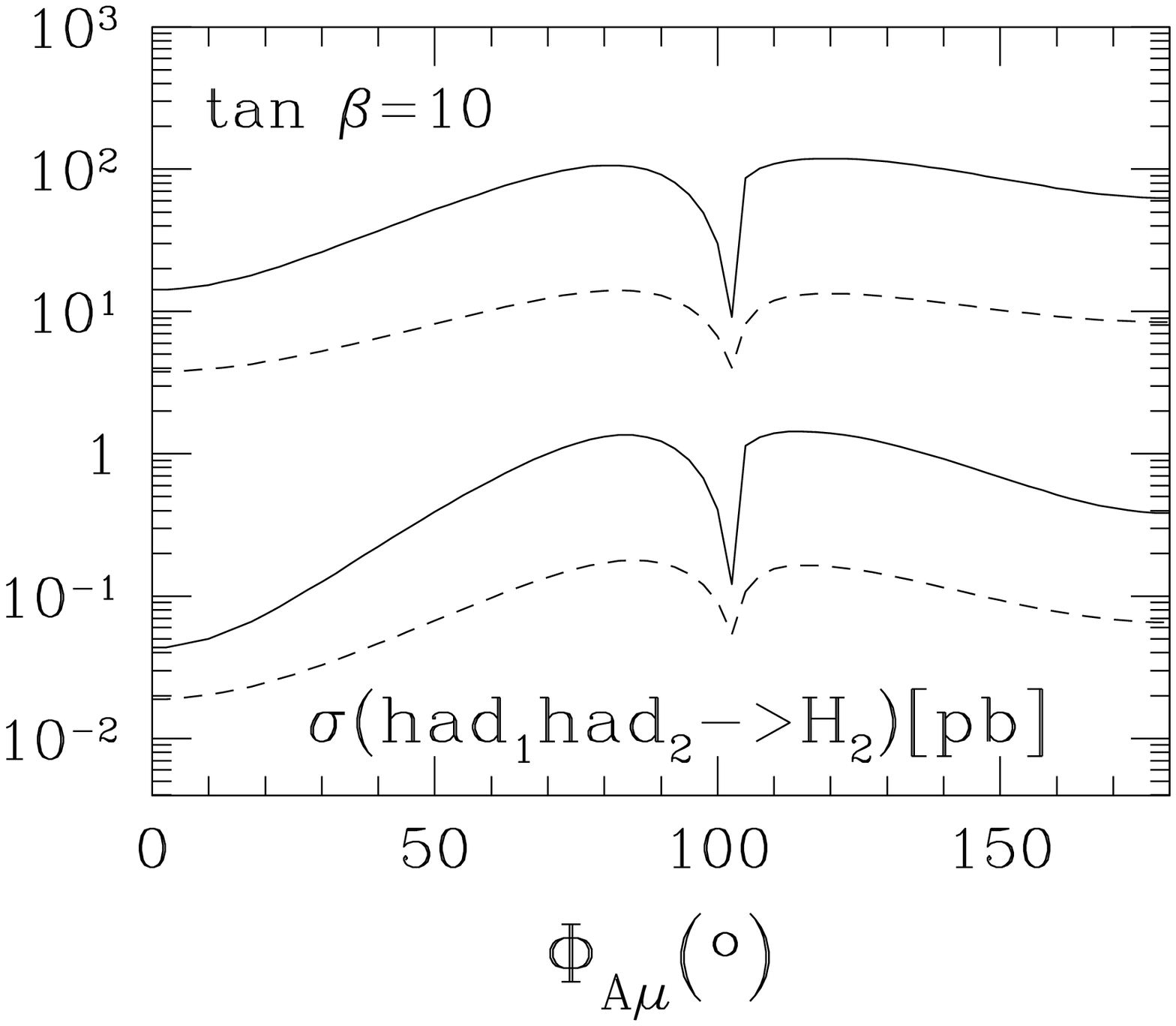}
\hspace{5mm}
\includegraphics[width=7.3cm]{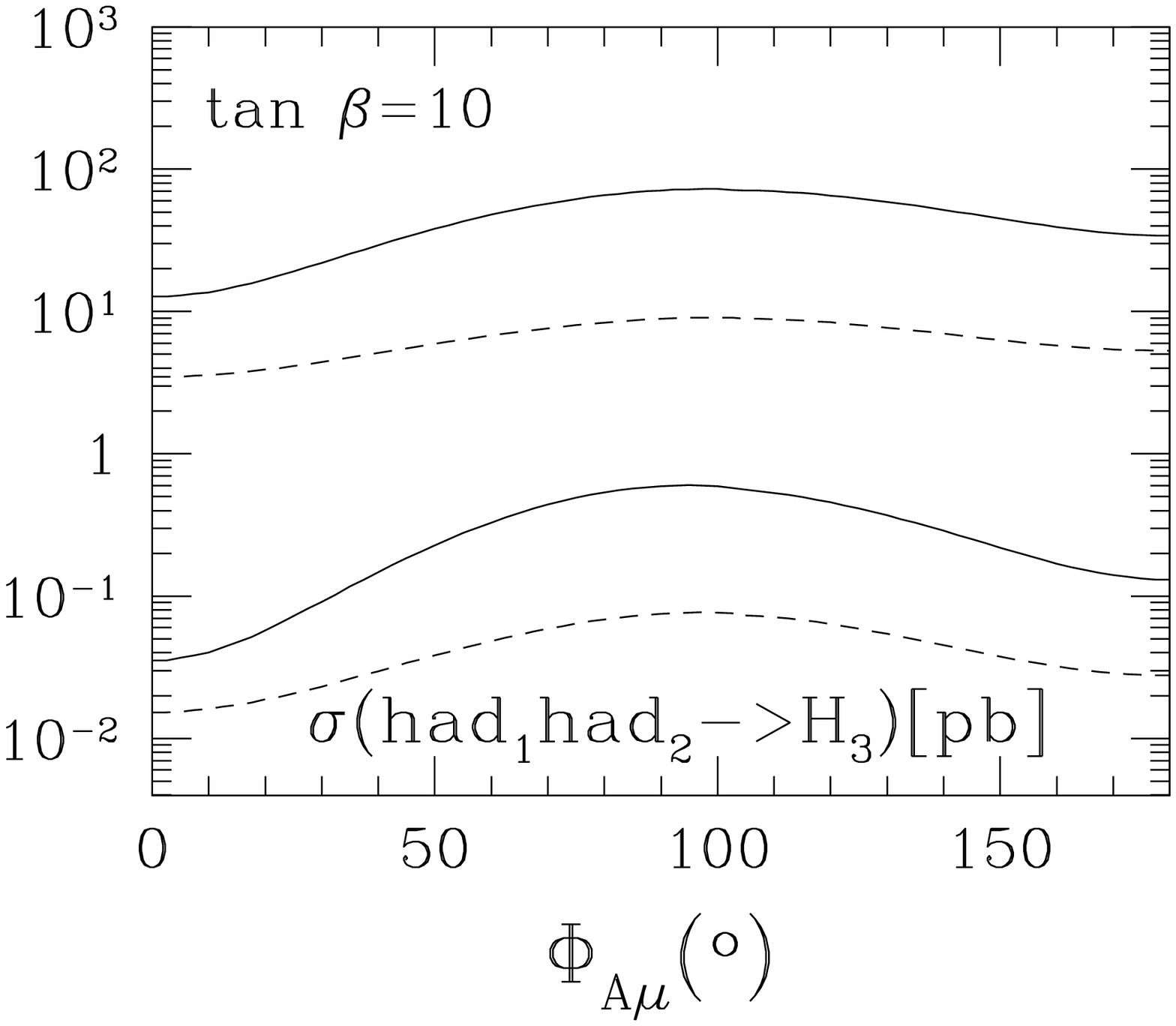}
\end{center}
\vspace{-0.4cm}
\caption{{\small \it Cross sections for the production of $H_1$,
 $H_2$, and $H_3$ vs. $\Phi_{A\mu}$, for
 $\Phi_{g\mu}=180^{\rm o}$~(solid lines) and
 $0^{\rm o}$~(dashed lines), at the LHC with $\sqrt{s}=14\,$TeV
 (two upper lines) and at the Tevatron with $\sqrt{s}=1.96\,$TeV 
 (two lower lines). ${\rm had}_1 {\rm had}_2 = pp$ for the LHC, 
  $p \bar{p}$ for the Tevatron.}}
\label{fig:Xsects}
\end{figure}

After convoluting the parton distribution functions, we obtain the
hadronic cross sections for the Tevatron ($\sqrt{s}=1.96\,$TeV) 
and the LHC ($\sqrt{s}=14\,$TeV). These are shown in 
Fig.~\ref{fig:Xsects} vs. 
$\Phi_{A\mu}$, for $\tan \beta =10$ and two values of 
$\Phi_{g\mu}$: $\Phi_{g\mu}=0^{\rm o}$~(dashed lines) and 
$\Phi_{g\mu}=180^{\rm o}$~(solid lines).  We have used the
leading-order CTEQ6L~\cite{CTEQ6} parton distribution functions and
chosen the factorization scale $Q=m_{H_i}/4$. This has been suggested
in most of the papers in Ref.~\cite{BBHsm} as the scale that minimizes
the next-to-leading-order QCD corrections to these cross sections when
no threshold corrections to $m_b$ are kept into account. Although this
should be explicitly checked, we believe that the inclusion of these
corrections, which amounts to substituting the tree-level Yukawa
couplings with effective ones, should not affect substantially this
result. We notice also that these supersymmetric threshold corrections
capture the main part of all supersymmetric corrections to the
production cross sections of neutral Higgs bosons through $b$-quark
fusion. Other corrections, with a nontrivial dependence on the
momenta of the $H_i$ bosons are of decoupling nature, and therefore
subleading.  (See the explicit check for the related processes $g b
\to b H_i$ in the CP-conserving case of Ref.~\cite{JMYang}.)

As already observed at the partonic level, all three production cross
section can deviate considerably from those obtained in CP-conserving
scenarios and the impact of the supersymmetric threshold corrections
to $m_b$ is very important.  The production cross section for the
lightest Higgs boson through the $b$-quark fusion is, in general,
comparable to the production cross section via gluon fusion. For some
values of $\Phi_{g\mu}$ and $\Phi_{A\mu}$, it can even be larger. For
example, for $\Phi_{g\mu} =\Phi_{A\mu}\sim 100^{\rm o}$, a production
cross section via gluon fusion of $\lsim 30\,$pb at the LHC, which can
be easily evinced from Ref.~\cite{CHL} for the same set of
supersymmetric masses used here, in particular the same value of
$m_{H_1}$, is indeed smaller than the corresponding cross sections in
Fig.~\ref{fig:Xsects}.  (The analysis of Ref.~\cite{CHL} does not
include the supersymmetric threshold corrections to $m_b$. Their
impact, however, is expected to be less dramatic in the case of the
gluon-fusion production.)  Even for different values of $\Phi_{g\mu}$
and $\Phi_{A\mu}$, the $b$-quark fusion production mechanism cannot be
easily dismissed as subleading.  No comparison between the two
production mechanisms for the two heavier Higgs bosons is possible at
the moment, since in the existing studies~\cite{glfusionHeavyH} of
$H_2$ and $H_3$ production via gluon fusion different supersymmetric
spectra than those considered here are analyzed.  For all three mass
eigenstates $H_i$, however, the cross sections corresponding to the
two production mechanisms have rather different dependences on the
supersymmetric spectrum. The results shown here for the $b$-quark
fusion cross sections, for example, remain unchanged for values of
$M_{\rm SUSY}$ different from our representative value of $0.5\,$TeV,
provided the relative size of the different supersymmetric parameters
is not changed. The same is not true for the gluon-fusion cross
sections, which are sensitive to the absolute value of $M_{\rm SUSY}$.

More investigations, theoretical and experimental, are needed to
unravel all implications of the enhanced cross sections presented
here. A detail comparison of the yield of all neutral Higgs bosons of
supersymmetric scenarios through $b$-quark fusion, as opposite to the
yield through gluon fusion and through Higgs strahlung (which is
relevant for the Tevatron), will have to be performed. Moreover, all
subsequent decay modes for these Higgs bosons need to be studied to
allow unambiguous interpretations of possible Higgs bosons signals,
which hopefully will be detected at the Tevatron and/or the LHC. All
future analyses need to be carried out at the same level of precision,
i.e. incorporating the threshold corrections to $m_b$ that, as proven
in this paper, turn out to be very important in these scenarios, not
only for large but also for intermediate values of $\tan \beta$.

To summarize, we have studied the effects of threshold corrections to
the $b$-quark mass on the production of neutral Higgs bosons via
$b$-quark fusion, $b\bar{b}\rightarrow H_i$, in supersymmetric
scenarios with large CP-violation in the Higgs sector.  This is
assumed to be induced by explicit phases in supersymmetric and
supersymmetry-breaking parameters.  For these particular scenarios, we
have found that:
\\[1.5ex]
\noindent {\it i)}
large phases of the combination $M_{\tilde{g}}\mu$, i.e.  
$\Phi_{g\mu} = 180^{\rm o}\pm 30^{\rm o}$, for all values of
$\Phi_{A\mu}$, can drive the mass squared of the lightest
$\tilde{b}$ squark to negative values and, depending on the
supersymmetric spectrum, the $b$-quark Yukawa coupling to
nonperturbative ones;
\\[1.5ex]
\noindent {\it ii)}
large deviations in the behaviour of the $b$-quark fusion cross
sections, with respect to the same cross sections in CP-conserving
scenarios, are obtained for $\Phi_{A\mu} \sim 100^{\rm o}$ when the
Higgs mixing is maximal;
\\[1.5ex]
\noindent {\it iii)}
the supersymmetric corrections to the $b$-quark fusion production
cross sections are in general very large, even for the four
CP-conserving scenarios obtained by fixing the values of $\Phi_{g\mu}$
and $\Phi_{A\mu}$ at $0^{\rm o}$ or $180^{\rm o}$. Among these four 
scenarios, those with $\Phi_{g\mu} =\Phi_{A\mu}=180^{\rm o}$ give the 
largest cross sections 
$\sigma({\rm had}_1 {\rm had}_2 \to b \bar{b} \to H_i)$ both at the
Tevatron and at the LHC.
\\[1.5ex]
We conclude that in CP-violating supersymmetric scenarios, the
production of neutral Higgs bosons through $b$-quark fusion cannot
be neglected, in general, with respect to the production through gluon
fusion.  Dedicated theoretical and experimental studies should be
carried out in order to allow a reliable detection of these mixed
state neutral Higgs bosons, or to constrain the CP-violating scenarios
that predict them.
%

\subsection*{Acknowledgements}

Discussions with A.~Pilaftsis and A.~Pomarol are acknowledged.
F.B. was partially supported by the Japanese Society for
Promotion of Science; J.S.L. by the Japanese Society for
Promotion of Science and PPARC; W.Y.S. by the 
KRF PBRG 2002-070-C00022 and KRF Grant 2000-015-DP0080. 
The hospitality of the theory groups of KEK, the University of
Barcelona, and the Yokohama National University is acknowledged.

{\small

}

\begin{thebibliography}{99}

\bibitem{bMASScorr}
T.~Banks,
Nucl.\ Phys.\ B {\bf 303} (1988) 172;
%
R.~Hempfling,
Phys.\ Rev.\ D {\bf 49} (1994) 6168;
%
L.~J.~Hall, R.~Rattazzi and U.~Sarid,
Phys.\ Rev.\ D {\bf 50} (1994) 7048;
%
T.~Blazek, S.~Raby and S.~Pokorski,
Phys.\ Rev.\ D {\bf 52} (1995) 4151;
%
M.~Carena, M.~Olechowski, S.~Pokorski and C.~E.~Wagner,
Nucl.\ Phys.\ B {\bf 426} (1994) 269;
%
D.~M.~Pierce, J.~A.~Bagger, K.~T.~Matchev and R.~j.~Zhang,
Nucl.\ Phys.\ B {\bf 491} (1997) 3;
%
K.~S.~Babu and C.~F.~Kolda,
Phys.\ Lett.\ B {\bf 451} (1999) 77;
%
F.~Borzumati, G.~R.~Farrar, N.~Polonsky and S.~Thomas,
Nucl.\ Phys.\ B {\bf 555} (1999) 53,
and
arXiv:hep-ph/9805314;
%
H.~Eberl, K.~Hidaka, S.~Kraml, W.~Majerotto and Y.~Yamada,
Phys.\ Rev.\ D {\bf 62} (2000) 055006;
%
H.~E.~Haber, M.~J.~Herrero, H.~E.~Logan, S.~Penaranda,
 S.~Rigolin and D.~Temes,
Phys.\ Rev.\ D {\bf 63} (2001) 055004.
%
For recent discussions, see also 
F.~Borzumati, C.~Greub and Y.~Yamada,
arXiv:hep-ph/0305063
and
Phys.\ Rev.\ D {\bf 69} (2004) 055005.


\bibitem{DEMIR}
D.~A.~Demir,
Phys.\ Lett.\ B {\bf 571} (2003) 193.


\bibitem{CPVHiggs0}
A.~Pilaftsis,
Phys.\ Rev.\ D {\bf 58} (1998) 096010, 
and 
Phys.\ Lett.\ B {\bf 435} (1998) 88.


\bibitem{CPVHiggs}
D.~A.~Demir,
Phys.\ Rev.\ D {\bf 60} (1999) 095007;
%
A.~Pilaftsis and C.~E.~Wagner,
Nucl.\ Phys.\ B {\bf 553} (1999) 3;
%
M.~Carena, J.~R.~Ellis, A.~Pilaftsis and C.~E.~Wagner,
Nucl.\ Phys.\ B {\bf 586} (2000) 92;
%
S.~Y.~Choi, M.~Drees and J.~S.~Lee,
Phys.\ Lett.\ B {\bf 481} (2000) 57.


\bibitem{REVIEWS}
H.~E.~Haber and G.~L.~Kane,
Phys.\ Rept.\  {\bf 117} (1985) 75;
%
K. Hikasa, Lecture Notes {\it ``Supersymmetric Standard Model for 
 Collider Physicists''} (1996) (unpublished);
%
A.~Djouadi {\it et al.}  [MSSM Working Group Collaboration],
arXiv:hep-ph/9901246;
%
N.~Polonsky,
Lect.\ Notes Phys.\  {\bf M68} (2001) 1,
arXiv:hep-ph/0108236.


\bibitem{Higgsrev}
M.~Carena {\it et al.},
arXiv:hep-ph/0010338;
%
D.~Cavalli {\it et al.},
arXiv:hep-ph/0203056;
%
M.~Carena and H.~E.~Haber,
Prog.\ Part.\ Nucl.\ Phys.\  {\bf 50}, 63 (2003).


\bibitem{EXP-CP-L} 
A.~Pilaftsis,
Phys.\ Rev.\ Lett.\  {\bf 77} (1996) 4996;
%
K.~S.~Babu, C.~F.~Kolda, J.~March-Russell and F.~Wilczek,
Phys.\ Rev.\ D {\bf 59} (1999) 016004;
%
S.~Y.~Choi and M.~Drees,
Phys.\ Rev.\ Lett.\  {\bf 81} (1998) 5509;
%
C.~A.~Boe, O.~M.~Ogreid, P.~Osland and J.~z.~Zhang,
Eur.\ Phys.\ J.\ C {\bf 9} (1999) 413;
%
B.~Grzadkowski, J.~F.~Gunion and J.~Kalinowski,
Phys.\ Rev.\ D {\bf 60} (1999) 075011;
%
S.~Y.~Choi and J.~S.~Lee,
Phys.\ Rev.\ D {\bf 61} (2000) 111702;
%
S.~Y.~Choi and J.~S.~Lee,
Phys.\ Rev.\ D {\bf 62} (2000) 036005;
%
S.~Bae,
Phys.\ Lett.\ B {\bf 489} (2000) 171;
%
E.~Asakawa, S.~Y.~Choi and J.~S.~Lee,
Phys.\ Rev.\ D {\bf 63} (2001) 015012;
%
E.~Asakawa, S.~Y.~Choi, K.~Hagiwara and J.~S.~Lee,
Phys.\ Rev.\ D {\bf 62} (2000) 115005;
%
M.~Carena, J.~R.~Ellis, A.~Pilaftsis and C.~E.~Wagner,
Phys.\ Lett.\ B {\bf 495} (2000) 155;
%
S.~Y.~Choi, M.~Drees, B.~Gaissmaier and J.~S.~Lee,
Phys.\ Rev.\ D {\bf 64} (2001) 095009;
%
M.~S.~Berger,
Phys.\ Rev.\ Lett.\  {\bf 87} (2001) 131801;
%
A.~G.~Akeroyd and A.~Arhrib,
Phys.\ Rev.\ D {\bf 64} (2001) 095018;
%
C.~Blochinger {\it et al.},
arXiv:hep-ph/0202199.


\bibitem{EXP-CP-H} 
J.~F.~Gunion and J.~Pliszka,
Phys.\ Lett.\ B {\bf 444} (1998) 136;
%
A.~Dedes and S.~Moretti,
Phys.\ Rev.\ Lett.\  {\bf 84} (2000) 22, 
and
Nucl.\ Phys.\ B {\bf 576} (2000) 29;
%
S.~Mrenna, G.~L.~Kane and L.~T.~Wang,
Phys.\ Lett.\ B {\bf 483} (2000) 175;
%
S.~Y.~Choi and J.~S.~Lee,
Phys.\ Rev.\ D {\bf 61} (2000) 115002;
%
S.~Y.~Choi, K.~Hagiwara and J.~S.~Lee,
Phys.\ Lett.\ B {\bf 529} (2002) 212;
%
M.~Carena, J.~R.~Ellis, S.~Mrenna, A.~Pilaftsis and C.~E.~M.~Wagner,
Nucl.\ Phys.\ B {\bf 659} (2003) 145;
%
B.~E.~Cox, J.~R.~Forshaw, J.~S.~Lee, J.~Monk and A.~Pilaftsis,
Phys.\ Rev.\ D {\bf 68} (2003) 075004.


\bibitem{EXP-CP-OTHER} 
S.~Y.~Choi and J.~S.~Lee,
Phys.\ Rev.\ D {\bf 61} (2000) 015003;
%
S.~Y.~Choi, K.~Hagiwara and J.~S.~Lee,
Phys.\ Rev.\ D {\bf 64} (2001) 032004;
%
S.~Y.~Choi, M.~Drees, J.~S.~Lee and J.~Song,
Eur.\ Phys.\ J.\ C {\bf 25} (2002) 307;
%
A.~Bartl, S.~Hesselbach, K.~Hidaka, T.~Kernreiter and W.~Porod,
Phys.\ Lett.\ B {\bf 573} (2003) 153.


\bibitem{CPX}
M.~Carena, J.~R.~Ellis, A.~Pilaftsis and C.~E.~Wagner,
in Ref.~\cite{EXP-CP-L}. 


\bibitem{EDMcontrib}
D.~Chang, W.~Y.~Keung and A.~Pilaftsis,
Phys.\ Rev.\ Lett.\  {\bf 82} (1999) 900
[Erratum-ibid.\  {\bf 83} (1999) 3972],
and references therein. 

 
\bibitem{EDMlargephases}
Y.~Kizukuri and N.~Oshimo,
Phys.\ Rev.\ D {\bf 45} (1992) 1806;
%
T.~Ibrahim and P.~Nath,
Phys.\ Rev.\ D {\bf 57} (1998) 478
[Errata-ibid.\ D {\bf 58} (1998) 019901, 
 \ D {\bf 60} (1999) 079903, \ D {\bf 60} (1999) 119901].


\bibitem{EDMandCOUPLINGS} 
A.~Pilaftsis,
Nucl.\ Phys.\ B {\bf 644} (2002) 263;


\bibitem{EDMconstr}
S.~Abel, S.~Khalil and O.~Lebedev,
Nucl.\ Phys.\ B {\bf 606} (2001) 151;
%
D.~Demir, O.~Lebedev, K.~A.~Olive, M.~Pospelov and A.~Ritz,
Nucl.\ Phys.\ B {\bf 680} (2004) 339.


\bibitem{CPMSSM}
M.~Dugan, B.~Grinstein and L.~J.~Hall,
Nucl.\ Phys.\ B {\bf 255} (1985) 413.
%
For a more modern formulation, see 
S.~Dimopoulos and S.~Thomas,
Nucl.\ Phys.\ B {\bf 465} (1996) 23.


\bibitem{CPsuperH} 
J.~S.~Lee, A.~Pilaftsis, M.~Carena, S.~Y.~Choi, M.~Drees, J.~Ellis and
C.~E.~Wagner,
Comput.\ Phys.\ Commun.\  {\bf 156} (2004) 283.


\bibitem{BFPT}
F.~Borzumati, G.~R.~Farrar, N.~Polonsky and S.~Thomas
in Ref.~\cite{bMASScorr}.


\bibitem{IbNa}
T.~Ibrahim and P.~Nath,
Phys.\ Rev.\ D {\bf 67} (2003) 095003
[Erratum-ibid.\ D {\bf 68} (2003) 019901],
%
and 
arXiv:hep-ph/0308167.


\bibitem{LEP2}
 LEP Higgs Working Group, CERN-EP/2003-011.


\bibitem{CHL}
S.~Y.~Choi, K.~Hagiwara and J.~S.~Lee, 
in Ref.~\cite{EXP-CP-H}. 


\bibitem{BOP}
F.~M.~Borzumati, M.~Olechowski and S.~Pokorski,
Phys.\ Lett.\ B {\bf 349} (1995) 311;
%
H.~Murayama, M.~Olechowski and S.~Pokorski,
Phys.\ Lett.\ B {\bf 371} (1996) 57;
%
R.~Rattazzi and U.~Sarid,
Nucl.\ Phys.\ B {\bf 501} (1997) 297;
%
F.~M.~Borzumati,
arXiv:hep-ph/9702307.


\bibitem{CTEQ6} 
J.~Pumplin, D.~R.~Stump, J.~Huston, H.~L.~Lai, P.~Nadolsky and 
W.~K.~Tung,
JHEP {\bf 0207} (2002) 012.


\bibitem{BBHsm} 
D.~A.~Dicus and S.~Willenbrock,
Phys.\ Rev.\ D {\bf 39} (1989) 751;
%
D.~Dicus, T.~Stelzer, Z.~Sullivan and S.~Willenbrock,
Phys.\ Rev.\ D {\bf 59} (1999) 094016;
%
C.~Balazs, H.~J.~He and C.~P.~Yuan,
Phys.\ Rev.\ D {\bf 60} (1999) 114001;
%
J.~Campbell, R.~K.~Ellis, F.~Maltoni and S.~Willenbrock,
Phys.\ Rev.\ D {\bf 67}, 095002 (2003);
%
F.~Maltoni, Z.~Sullivan and S.~Willenbrock,
Phys.\ Rev.\ D {\bf 67}, 093005 (2003);
%
E.~Boos and T.~Plehn,
arXiv:hep-ph/0304034;
%
R.~V.~Harlander and W.~B.~Kilgore,
Phys.\ Rev.\ D {\bf 68}, 013001 (2003);
%
S.~Dittmaier, M.~Kramer and M.~Spira,
arXiv:hep-ph/0309204;
%
S.~Dawson, C.~B.~Jackson, L.~Reina and D.~Wackeroth,
arXiv:hep-ph/0311067.


\bibitem{JMYang}
J.~j.~Cao, G.~p.~Gao, R.~J.~Oakes and J.~M.~Yang,
Phys.\ Rev.\ D {\bf 68} (2003) 075012;
%
H.~S.~Hou, W.~G.~Ma, R.~Y.~Zhang, Y.~B.~Sun and P.~Wu,
JHEP {\bf 0309} (2003) 074.


\bibitem{glfusionHeavyH}
A.~Dedes and S.~Moretti, in Ref.~\cite{EXP-CP-H}; 
S.~Y.~Choi and J.~S.~Lee, in Ref.~\cite{EXP-CP-H}. 


\end{thebibliography}
\end{document}